\documentclass[aps,jmp]{revtex4}
\usepackage{amsmath}
\usepackage{graphicx}
\usepackage{amssymb}
\begin{document}
 \def\g{\mbox{${\bf g}$}}
 \def\F{\mbox{${\cal F}$}}
\newcounter{rown}
\def\bl{\setcounter{rown}{\value{equation}}
        \stepcounter{rown}\setcounter{equation}0
        \def\theequation{\thesection.\arabic{rown}\alph{equation}}
        }
\def\el{\setcounter{equation}{\value{rown}}
        \def\theequation{\thesection.\arabic{equation}}
        }
\def\sec{\setcounter{equation}0}
\renewcommand{\theequation}{\thesection.\arabic{equation}}

\title{Basic Twist Quantization \\ of the Exceptional Lie Algebra $g_2$}

\author{Andrzej Borowiec}\thanks{Supported by KBN grant1PO3BO1828}
\affiliation{Institute for Theoretical Physics, \\ University of Wroc\l{}aw, \\
             pl. M. Borna 9, 50-204 Wroc{\l}aw, Poland}

 \author{Jerzy Lukierski}\thanks{Supported by KBN grant1PO3BO1828}
\affiliation{Institute for Theoretical Physics, \\ University of Wroc\l{}aw, \\
             pl. M. Borna 9, 50-204 Wroc{\l}aw, Poland}

\author{Vladimir~Lyakhovsky}\thanks{on leave of absence from Sankt-Petersburg University,
         Ulianovskaya 1, Petrodvoretz, 198904 Sankt Petersburg, Russia}

\affiliation{Institute for Theoretical Physics, \\ University of Wroc\l{}aw, \\
             pl. M. Borna 9, 50-204 Wroc{\l}aw, Poland}

\author{
 Marek Mozrzymas}\thanks{Supported by KBN grant1PO3BO1828}
\affiliation{Institute for Theoretical Physics, \\ University of Wroc\l{}aw, \\
             pl. M. Borna 9, 50-204 Wroc{\l}aw, Poland}

\author{Valeriy  N. Tolstoy}\thanks{on leave of absence from Institute of Nuclear Physics, Moscow State University, 119 992 Moscow, Russia}
\affiliation{Institute for Theoretical Physics, \\ University of Wroc\l{}aw, \\
             pl. M. Borna 9, 50-204 Wroc{\l}aw, Poland}

\date{\today}

\begin{abstract}
We present the formulae for twist quantization of $g_2$,
corresponding to the solution of classical YB equation with
support in the 8-dimensional Borel subalgebra of $g_2$. The
considered chain of twists consists of the four factors describing
the four steps of quantization: Jordanian twist, the two twist
factors extending Jordanian twist and the deformed Jordanian or in
second variant additional Abelian twist. The first two steps describe
as well the $sl(3)$ quantization. The coproducts are calculated
for each step in explicite form, and  for that purpose we present
new formulas for the calculation of similarity transformations on
tensor product. We introduce new basic generators in universal
enveloping algebra $U(g_2)$ which provide nonlinearities in
algebraic sector maximally simplifying the deformed coproducts.
\end{abstract}

\pacs{Valid PACS appear here}% PACS, the Physics and Astronomy
\maketitle

\section{Introduction}
In this paper we shall consider the basic
 nonstandard quantum deformations of complex exceptional Lie
 algebra
 $g_2$.  %\footnote{We denote
%   by $G_2$ the Lie algebra; the corresponding group will be
%    called explicitly ``$G_2$ group''.}.
  There are four complex
 semisimple Lie
 algebras of rank 2, given
  by $A_2 \simeq  sl(3)$, $D_2 \simeq o(4)=o(3) \oplus o(3)$,
   $B_2 \simeq C_2 \simeq o(5) \simeq
 sp(4)$ and $g_2$,  with 8, 6, 10 and 14 generators respectively.
The 8-dimensional  carrier of classical $r$-matrices which describe
our deformations is equal to the  Borel subalgebra
${\mathfrak{b}}_+ (g_2)$ of   $g_2$.

 There are two natural embeddings related with the group $G_2$:

 i) $G_{2} \subset O(7) \, $.
  The fundamental matrix representation of $G_2$ is seven-dimensional.
 The $7\times 7$ orthogonal matrices $(L_{\rm ab})
 \in O(7)$ (${\rm a,b}=1,2, \ldots , 7$)
  belong
 to the group
   $G_{2}$ if the following cubic constraint is satisfied \cite{1basic,4basic}

\begin{equation}\label{blm1}
  f_{\rm a_1 a_2 a_3} = f_{\rm b_1 b_2 b_3} \, L_{\rm a_1 b_1} \,
  L_{\rm a_2 b_2} \,L_{\rm a_3 b_3} \, ,
\end{equation}
where  the totally antisymmetric cubic tensor $f_{\rm abc}$ describes the
multiplication table for imaginary octonions $t_a$
\begin{equation}\label{blm2}
  t_{\rm a} t_{\rm b} = f_{\rm abc} t_{\rm c}
\end{equation}
and the values of $f_{\rm abc}$ are determined by the following
choice (we list only nonvanishing values)
\begin{equation}\label{blm3}
  f_{\rm 127} = f_{\rm 157} = f_{\rm 163} = f_{\rm 264} = f_{\rm
  245}= f_{\rm 374} = f_{\rm 576} = 1\, .
\end{equation}
Therefore there are only seven independent equations (\ref{blm1})
reducing 21 parameter of $O(7)$ to 14 parameters of the group $G_2$.

We see that the fundamental seven-dimensional representation $\{ 7\}$ of
Lie algebra $g_2$ inherits basic properties of the fundamental $o(7)$
representation: reality and  its dimensionality.
 The generators  ($E_k^{\ l}, A_k, B^l$) ($k,l=1,2,3$) of  $g_2$
satisfy the following  relations  (\cite{1basic}--\cite{14basic})
%\cite{14basic})
 \bl
\begin{eqnarray}
%\beeqa
\label{1.4a}
[E_k^{\ l}, E_m^{\ n}] & = & \delta_m^{\ l} E_k^{\ n} -
\delta_k^{\ n} E_m^{\ l} \, , \\
\cr
\label{1.4b}
[E_k^{\ l}, A_m ] & = & \delta_m^{\ l} A_{ k} -
\frac{1}{3} \delta_k^{\ l} A_m \, , \\
\cr
\label{1.4c}
[E_k^{\ l}, B^n] &  =&  - \delta_k^{\ n} B^{\ l} +
\frac{1} {3} \delta_k^{\ l} B^{ n} \, , \\
\cr
\label{1.4d}
[A_m, B^n] & = & E_m ^{\ n}\, , \\
\cr
\label{1.4e}
[A_m, A_n] & =& -  \frac{4}{{3}} \, \epsilon_{\rm mnl}\, B^l\, , \\
\cr
\label{1.4f}
[B^m, B^n] & = &
   \epsilon^{\rm mnl}\, A_l\, ,
\end{eqnarray}
where
\begin{equation}\label{1.4g}
E^{\ k}_{ k} = 0 \, ,
\end{equation}
 \el
 and we employ the $su(3)$ tensorial basis.
%%%%%%%%

ii) $sl(3) \subset g_2$ or $su(3) \subset g_2$ (real form).
 The generators $E_k^{\ l}$ forming the subalgebra (\ref{1.4a})
describe $sl(3)$ (if $g_2$ is complex Lie algebra) or
 {$su(3)$ (if we introduce in $g_2$ the  suitable real structure).}
The Lie algebra generators of $g_2$
  belong to 14-dimensional adjoint representation
$\{ \underline{14} \}$ which decomposes under $sl(3)$ (or $su(3)$) as follows:
\begin{equation}\label{bml1.5}
  \{ \underline{14} \} = \{ \underline{8} \} + \{ \underline{3} \}
   + \{ \underline {\overline{3}} \} \, .
\end{equation}
In the realization (\ref{1.4a}--\ref{1.4f}) of the Lie algebra
 $g_2$
   the generators $A_m$ ($B^m$) transform as fundamental triplet (antitriplet)
representations of $sl(3)$ or $su(3)$. These properties can be
also seen from the root diagram of $g_2$ (see Sect.~II).

The relations (\ref{1.4d}--\ref{1.4g})
  show that the coset space $S^6 = \frac{G_2}{SU(3)}$
is a nonsymmetric Riemannian space, with torsion described by nonvanishing rhs of relations
(\ref{1.4e}--\ref{1.4f}).
One of the aims of this paper is to provide an algebraic ground for new
quantum deformation of the  sphere $S^6$ with torsion.

The embedding of 3-dimensional fundamental representation of $su(3)$
described by Gell-Mann fundamental matrices $\lambda_k$ ($k=1,\ldots 8$) into
 7-dimensional fundamental representation of $g_2$ looks as follows
\begin{equation}\label{blm1.6}
\Lambda_k = \frac{1}{\sqrt{2}}
  \begin{pmatrix}
%{
  \lambda_{k} & 0 & 0 \cr
    0  & - \lambda_{k}^*  & 0 \cr
    0 &  0  &  0
    % }
  \end{pmatrix}\, ,
\end{equation}
where we use the standard normalization
\begin{equation}\label{bml1.7}
  Tr\, \lambda_k \, \lambda_l = Tr     \, \Lambda_k \, \Lambda_l = 2 \delta_{kl}\, .
\end{equation}
Two Cartan generators of $g_2$
 we identify with
 the $su(3)$ generators $\Lambda_3$ and $\Lambda_8$.

%%%%%%

       The plan of our paper is the following:

       In Sect.~II we shall consider the Lie algebra $g_2$ in Cartan-Weyl basis
       (see e.g. \cite{5basic}) which is directly linked with   the generators
       ($E_{k}^{\ l}, A_k, B^l$) satisfying the algebra (\ref{1.4a}--\ref{1.4g}).
       We  present the important class of triangular  $r$-matrices for
        $g_2$, satisfying
        the  classical Yang-Baxter equation (CYBE).
           It appears that the
         two-parameter  families of such
         $r$-matrices have as its carrier algebra the
     whole     8-dimensional Borel subalgebra
     ${\mathfrak{b}}_+(g_2) \subset g_2$.
     We show that the  parameters of the considered classical r-matrices
      can achieve fixed nonzero values by means of inner automorphism maps
       inside $g_2$ algebra.
         In Sect.~III we shall recall the general formulae which
           describe the twist quantization method \cite{6basic}--\cite{8basic},
             and  we shall introduce the general  twisting function, describing the
  twist quantization procedure for  $g_2$ with the 8-dimensional
   carrier space
    for its $r$-matrix. In Sect.~IV  we
 obtain
 firstly the explicite formulae  describing  the twist quantization of $g_2$
 generated by the $sl(3)$ classical $r$-matrix.
 It appears that these quantization formulae for  $g_2$ describe the  extension
 of known relations describing the twist quantization of $sl(3)$
  \cite{8basic}. In particular following general technique presented in
 \cite{9basic} we shall introduce suitable nonlinear basis in the deformed
  Hopf algebras.
 In Sect.~V we consider the most general  $g_2$-quantizations
  containing
  two additional twists, depending on the $g_2$ generators from the coset
 $\frac{G_2}{sl(3)}$. In Sect.~VI we present
  a general discussion and
  some conclusions.
We remark that in Sect.~IV and V we shall use new algebraic
formulae for calculating twisted coproducts presented in Sect~III.D and shall introduce
 new basis of $U(g_2)$
 which will simplify the twisted coproduct formulae.
%%%

The motivation for our work is mainly to present a new mathematical result -
 interesting class of quantum deformations for an important Lie algebra. On the other side
  it should be stressed that  $g_2$ algebra recently has attracted attention
 of physicists
   in the
  domain of elementary particle physics and fundamental interactions theory.
  In particular we recall that:

  i) In eleven-dimensional $M$-theory there were proposed the internal
  manifolds with $g_2$ holonomy as a base for the grand unification describing extension
   of the standard model in particle physics
    (see e.g. \cite{10basic}--\cite{13basic}). The algebra $g_2$ implies
    seven-dimensional internal symmetry space as the privileged one,  in obvious connection with
    the relation $11 = 4+7$.

    ii) In the reduction of supersymmetric theories from $D=11$  to $D=4$ the $g_2$
    internal symmetry implies phenomenologically interesting case of $D=4$ models
     with $N=1$ supersymmetry \cite{14basic}. In particular there were also considered standard
      and supersymmetric   extensions of $D=4$ chromodynamics to $G_2$ gauge
      theories \cite{14basic} with interesting exceptional quark confinement mechanism.

      iii) There are four Hurwitz algebras (real numbers $R$, complex numbers $C$,
      quaternions $H$ and octonions $O$); $G_2$ acts on seven imaginary octonionic
       units and describes the automorphism group of the octonion algebra.
       All applications of exceptional and octonions groups to the description of
       symmetries in elementary particle physics
        (see e.g. \cite{15basic}) is strongly linked therefore with   the
       appearance of $G_2$ symmetry.

 In this paper we consider only the quantum
   deformations of universal enveloping  algebra
  $U(g_2)$; it is an interesting problem to supplement the considerations with
   deformations of dual Hopf algebra describing
    matrix quantum  $G_2$ group and further describe e.g. the quantum
    deformations of $S^6 = \frac{G_2}{SL(3)}$. In such a way one can obtain an example of
    six-dimensional counterpart of two-dimensional  Podle\'{s} sphere
     \cite{16basic},
     provided by  the deformed coset
     $\frac{SU(2)}{U(1)}$.

\section{Cartan-Weyl Basis of $g_2$ and Jordanian Type Classical
 $r$-matrices}

\subsection{Cartan-Weyl Basis of $g_2$}

\setcounter{equation}{0}
In order to describe Cartan--Weyl basis of $g_2$ let us introduce the
Dynkin diagram for its simple roots
   $\Pi=\{\alpha_1^{},\alpha_2^{}\}$:

%\begin{figure}[h]
%%{\thicklines
%%%%%%%%%%%%%%%%%%%%%%%%%%%%%\vskip 20pt
%%%%%%%%%%%%%%%%%%%%%%%%%%%%%%%%%%%%%%%
%\begin{center}
\begin{eqnarray*}
%%\begin{picture}(200,20)
%%\put(153,9){\line(1,0){32}}
%%\put(149.5,6){\circle{9}}
%%\put(191.5,6){\circle{9}}
%%\put(183,13){\line(1,-2){5}}
%%\put(183,-1){\line(1,2){5}}
%%\put(154,6){\line(1,0){33}}
%%\put(153,3){\line(1,0){32}}
%%\put(147,19){$\alpha_1^{}$}
%%\put(187,19){$\alpha_2^{}$}
%%\end{picture}%%}
%\end{eqnarray*}
%\begin{center}
\begin{picture}(100,20)
\put(53,9){\line(1,0){32}}
%\put(54,7){\line(1,0){32}}
\put(49.5,6){\circle{9}}
\put(91.5,6){\circle{9}}
\put(83,13){\line(1,-2){5}}
\put(83,-1){\line(1,2){5}}
\put(54,6){\line(1,0){33}}
\put(53,3){\line(1,0){32}}
\put(47,19){$\alpha_1^{}$}
\put(87,19){$\alpha_2^{}$}
\end{picture}%%}
%\end{center}
\end{eqnarray*}
%\caption{Dynkin diagram of the Lie algebra $g_2$.}
%\end{figure}
%%%%%%%%%%%%%%%%%%%%%%%%%%%%%%%%%%%%%%%
\vskip 8pt \centerline{\footnotesize Fig.~1. Dynkin diagram of the Lie algebra $g_2$.}
 \vskip 12pt \noindent
 The corresponding
standard $A=(a_{ij})  {(i,j=1,2)}$ and symmetric
$A^{sym}=(a_{ij}^{sym})_{i,j}$ Cartan matrices  are given by
             \begin{equation}
\label{}
A=
\begin{pmatrix}
%\pmatrix{
 2{\ } \, \, &-1
\cr
-3&{\ }\,\, 2
%}
\end{pmatrix}
 ~, \qquad
A^{sym}=
%\pmatrix{
\begin{pmatrix}
{\ }\,\,6&-3\cr-3&{\ }\,\, 2
%}
\end{pmatrix}
 \, . \qquad
\label{cw1}
\end{equation}

The Lie algebra $g_2$ is generated by the six Chevalley elements
$e_{\alpha_i}^{}$, $e_{-\alpha_i}^{}$, $h_{\alpha_i}^{}$ ($i=1,2$)
with the defining relations (see e.g. \cite{5basic})
%%\bl
\begin{eqnarray}
\begin{array}{rcl}
[h_{\alpha_i}^{},h_{\alpha_j}^{}]&=&0\,,
\\[5pt]
%%\cr
[h_{\alpha_i}^{},e_{\pm\alpha_j}^{}]&=&\pm a_{ij}^{sym}\,e_{\pm\alpha_j}\,,
\phantom{aaaaaaa}
\\[5pt]
[e_{\alpha_i}^{},e_{-\alpha_j}^{}]&=&\delta_{ij}^{}\,h_{\alpha_i}^{} \, ,
\\[5pt]
[e_{\pm\alpha_1}^{},[e_{\pm\alpha_1}^{},e_{\pm\alpha_2}^{}]]&=&0\,,
\\[5pt]
[[[[e_{\pm\alpha_1}^{},e_{\pm\alpha_2}^{}],e_{\pm\alpha_2}^{}],
e_{\pm\alpha_2}^{}],e_{\pm\alpha_2}^{}]&=&0\,.
\label{cw2}
\end{array}
\end{eqnarray}
%%\el

The positive $\Sigma_{+}^{}(g_2)$ and total
$\Sigma(g_2)=\Sigma_{+}^{}(g_2)\bigcup\,(-\Sigma_{+}(g_2))$ root systems
of $g_2$ are presented in terms of an orthonormalized basis
$\{\epsilon_1^{},\,\epsilon_2^{}\}$ of a 2-dimensional Euclidian space as
follows
\begin{eqnarray}
\Sigma_{+}^{}(g_2)\!\!&=&\!\!\Bigl\{\sqrt{3}\epsilon_1^{},\,\epsilon_2^{},\,
\mbox{\Large$\frac{\sqrt{3}}{2}$}\epsilon_1^{}\pm\mbox{\Large$\frac{1}{2}$}
\epsilon_2^{},\,\mbox{\Large$\frac{\sqrt{3}}{2}$}\epsilon_1^{}
\pm\mbox{\Large$\frac{3}{2}$}\epsilon_2^{}\Bigr\}\, \label{cw3}
\\[7pt]
\Sigma(g_2)\!\!&=&\!\!\Bigl\{\pm\sqrt{3}\epsilon_1^{},\,\pm\epsilon_2^{},\,
\pm\mbox{\Large$\frac{\sqrt{3}}{2}$}\epsilon_1^{}
\pm\mbox{\Large$\frac{1}{2}$}\epsilon_2^{},\,
\pm\mbox{\Large$\frac{\sqrt{3}}{2}$}\epsilon_1^{}
\pm\mbox{\Large$\frac{3}{2}$}\epsilon_2^{}\Bigr\}\,. \label{cw4}
\end{eqnarray}
where the simple roots are given by
$\alpha_1^{}=\frac{\sqrt{3}}{2}\epsilon_1^{}-\frac{3}{2}\epsilon_2^{}$ and
$\alpha_2^{}=\epsilon_2^{}$.
It is convenient to present the total root system by the
   root diagram presented on Fig.~2.
\setcounter{figure}{1}
\begin{figure}[hb]
\begin{center}
\includegraphics[width=7.5cm,angle=0]{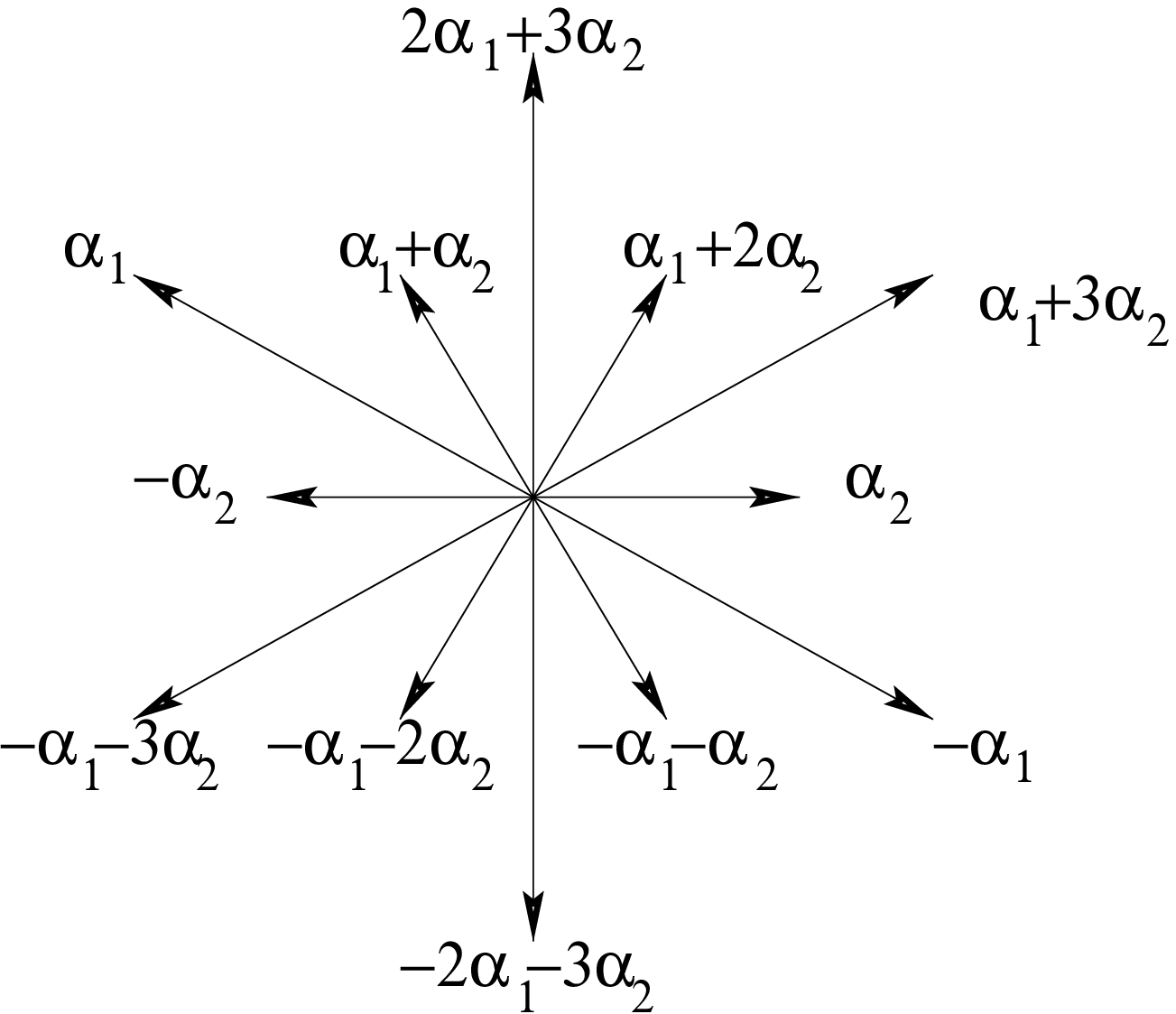}
%{\LARGE (I had not a file of this Figure)}
\end{center}
\caption{The root diagram for $g_2$} \label{wsp}
\end{figure}
%%%%%

For construction of the composite root vectors $e_{\gamma}^{}$,
($\gamma\neq\pm\alpha_1^{},\,\pm\alpha_2^{}$), we fix the
following normal ordering of the positive root system
$\Sigma_{+}^{}(g_2)$ (see \cite{5basic})
\begin{equation}
\alpha_{1}^{}\,,\alpha_{1}^{}+\alpha_{2}^{}\,,2\alpha_{1}^{}+3\alpha_{2}^{}\,,
\alpha_{1}^{}+2\alpha_{2}^{}\,,\alpha_{1}^{}+3\alpha_{2}^{}\,,\alpha_{2}^{}\,,
\label{cw5}
\end{equation}
which corresponds to "clockwise" ordering for positive roots in Fig.~2
if we start from the root $\alpha_{1}^{}$ to the root $\alpha_{2}^{}$.
For convenience  we introduce the short notations
\begin{equation}
e_{k,l}^{}:= e_{k\alpha_1+l\alpha_2}^{}\,,
\qquad
h_{k,l}^{}:=kh_{\alpha_1}^{}+lh_{\alpha_2}^{}
\label{cw6}
\end{equation}
for $k,l=0,\pm 1,\ldots$\,.
According to the ordering (\ref{cw5}) we set
 the composite roots generators with suitably chosen numerical
 coefficients as follows
\begin{equation}\label{bm2.5}
\begin{array}{ll}
%e_{ \alpha + \beta} =
e_{1, 1} =  [e_{1,0}, e_{0,1}]\, , \quad
&
% e_{ -\alpha - \beta} =
e_{-1, -1} = - [e_{- 1,0}, e_{0,-1}]\, ,
\cr
%e_{ \alpha + 2\beta} =
e_{1,2} =  [e_{1,1}, e_{0,1}]\, , \quad
&
%e_{ -\alpha - 2\beta} =
e_{-1,-2} = - \frac{3}{4} [e_{0,-1}, e_{-1,- 1}]\, ,
\cr
%e_{\alpha + 3\beta} =
e_{1,3} =  [e_{1,2}, e_{0,1}]\, ,\quad
&
%e_{-\alpha - 3\beta} =
e_{- 1,-3} = - \frac{3}{4} [e_{0,-1}, e_{-1,-2}]\, ,
\cr
%e_{2\alpha + 3\beta)} =
e_{2,3} =  [e_{1,3}, e_{1,0}]\, , \quad
 &
 %e_{-2\alpha - 3\beta)} =
 e_{-2,-3} = - \frac{3}{4} [e_{-1,0}, e_{-1,-3}]\, .
\end{array}
\end{equation}

The complete set of relations for Cartan-Weyl basis of $g_2$
 can be read off from  the formulae
(\ref{1.4a}--\ref{1.4f}) after the identification
\bl
\begin{equation}\label{bm2.9abis}
h_{1,0}\equiv  h_1 = E^{\ 2}_{ 2} - E^{\ 3} _{ 3} \, ,
  \qquad
h_{0,1} \equiv  h_2   =  \frac{1}{6}(E^{\ 1}_{ 1} - 2E^{\ 2}_{ 2}
 + E^{\ 3}_{3})
\end{equation}
and
\begin{equation}\label{mb2.9bis}
\begin{array}{rlllrl}
 e_{1,0}  &=&E^{\ 3}_{ 2} \, , \qquad
 &e_{-1,0} &= & E^{\ 2}_{ 3} \, ,
 \cr
 e_{0,1}  &=& B^2 \, , \qquad
  &e_{0,-1} &= & A_{2} \, ,
\cr
  e_{1,1}  &=& - B^3 \, , \qquad
  &e_{-1,-1} &= & - A_{ 3} \, ,
\cr
  e_{1,2}  &=& A_{1} \, , \qquad
  &  e_{-1,-2} &= &  B^1 \, ,
\cr
  e_{1,3}  &=& E^{\ 2}_{ 1} \, , \qquad
  & e_{-1,-3} &= &  E^{\ 1}_{ 2} \, ,
\cr
  e_{2,3}  &=& E^{\ 3}_{ 1} \, , \qquad
  & e_{-2,-3} &= &  E^{\ 1}_{ 3} \, .
\end{array}
\end{equation}
\el

\subsection{Jordanian Type Classical $r$-Matrices for $g_2$}

Firstly we introduce some definitions concerning classical $r$-matrices.
Let $\mathfrak{g}$ be any simple Lie algebra then $\mathfrak{g}=
\mathfrak{n_-^{}}\oplus\mathfrak{h}\oplus\mathfrak{n_+^{}}$ where
$\mathfrak{n_{\pm}^{}}$ are maximal nilpotent subalgebras and
$\mathfrak{h}$ is a Cartan subalgebra. The subalgebra $\mathfrak{n_+^{}}$
($\mathfrak{n_-^{}}$) is generated by the positive (negative) root vectors
$e_{\beta}^{}$ ($e_{-\beta}^{}$ ) for all
$\beta\in \Sigma_+({{\mathfrak{g}}})$.
The symbol $\mathfrak{b}_{+}^{}$ will denote the Borel subalgebra of
$\mathfrak{g}$, $\mathfrak{b}_{+}^{}:=\mathfrak{h}\oplus\mathfrak{n}_+^{}$.
Let the elements $h_{\theta}^{}\in \mathfrak{h}$ and
$e_{\theta}^{}\in \mathfrak{n}_{+}$ satisfy  the relation
\begin{equation}
[h_{\theta}^{},\,e_{\theta^{}}]=e_{\theta}^{}\,.
\label{cw11}
\end{equation}
A two-tensor $r_{J}^{}(\xi_{\theta}^{})\in \mathfrak{b}_+
\otimes\mathfrak{b}_{+}$ of the form
\begin{equation}
r_{\theta}^{}
(\xi_{\theta}^{})=\xi_{\theta}^{}\,h_{\theta}^{}\wedge
e_{\theta}^{}:=\xi_{\theta}^{}\bigl(h_{\theta}^{}\otimes e_{\theta}^{}-
e_{\theta}^{}\otimes h_{\theta}^{}\bigr)\,
\label{cw12}
\end{equation}
satisfies CYBE and it is called the Jordanian classical $r$-matrix.
The symbol $\xi_{\theta}^{}\in {\mathbb{C}}$ is a deformation parameter.
Moreover, let elements $e_{\gamma_{\pm i}}$, indexed by the symbols $i$ and $-i$,
$i\in I=\{1,2,\ldots,N\}$ satisfy the relations
\bl
\begin{eqnarray}
%\begin{array}{lrrll}
&[h_{\theta}^{},\,e_{\gamma_i}]=(1-t_{\gamma_i})\,e_{\gamma_i}\,,
\quad
&[h_{\theta}^{},\,e_{\gamma_{-i}}]=
t_{\gamma_i}\,e_{\gamma_{-i}}  \,,
\\ \cr
&[e_{\gamma_i},\,e_{-\gamma_j} ]=\delta_{ij}^{}\,e_{\gamma_0}\,
\quad\qquad
& [e_{\gamma_{ \pm i}}, e_{\gamma_{ \pm j}} ]= 0\, ,
%%%\
%% \bigr(k,\,l \in I\bigcup(-I)\bigl).
\\ \cr
&
[ e_{\gamma_{\pm i}} , e_{\gamma_0} ] = 0 \, . &
\label{cw13}
%\end{array}
\end{eqnarray}
\el
It is not difficult to check (see also \cite{7basic})  that
the element
\begin{equation}
r_{\theta;N}^{}
(\xi_{\theta}^{})=\xi_{\theta}^{}\Bigl(h_{\theta}^{}\wedge
e_{\theta}^{}+\sum_{i=1}^{N}e_{\gamma_i}\wedge e_{\gamma_{-i}}\Bigr)
\label{cw14}
\end{equation}
satisfies CYBE and it will be called the extended Jordanian $r$-matrix
of $N$-order. Let $N$ be maximal order, i.e. there do not exist
other elements $e_{\gamma_{\pm j}}^{}\in \mathfrak{n}_+^{}$, $j> N$,
which satisfy the relations (\ref{cw13}) then the element (\ref{cw14})
will be called the extended Jordanian $r$-matrix of maximal order
 \cite{drin20}.
It is evident that the extended Jordanian $r$-matrix of maximal order is
defined by the elements $h_{\theta}^{}\in\mathfrak{h}$,
$e_{\theta}^{}\in\mathfrak{n}_{+}$ and the Borel subalgebra
$\mathfrak{b}_+$. We shall here consider a special
("canonical") case when  $e_{\theta}^{}$ and $e_{\gamma_{\pm i}}$
($i=1,2,\cdots,N$) are weight elements with respect to the Cartan
subalgebra $\mathfrak{h}$
\begin{equation}
[h,\,e_{\theta}^{}]=(h,\theta)\,e_{\theta}^{}\,,\qquad
[h,\,e_{\gamma_{\pm i}}]=(h,\gamma_{\pm i})\,e_{\gamma_{\pm i}}\,
\label{cw15}
\end{equation}
for any $h\in\mathfrak{h}$ and for all $i=1,2,\cdots,N$. Analyzing
the structure of the positive root systems of the complex simple Lie
algebras we see that if
 $e_{\gamma_{\pm j}}^{}\in \mathfrak{n}_+^{}$
 the  maximal order $N$ of the extended
Jordanian $r$-matrix is associated with the maximal root, i.e. the root
$\theta$ is maximal.

Let us pass now to the Lie algebra $ \mathfrak{g}= g_2$. The
maximal root
 generator $e_\theta$ is $e_{2,3} = e_{2\alpha_1 + 3 \alpha_2}$ and the extended
Jordanian matrix of maximal order is provided by
formula (\ref{cw14}) with $N=2$. It takes the form:

\begin{equation}
r_{2,3;2} (\xi)
= \xi
\left(h_{2,3}\wedge
e_{2,3}
+ e_{1,1}\wedge
e_{1,2} + e_{1,3} \wedge e_{1,0}\right)\, .
\label{cw14bis}
\end{equation}

In order to obtain the generalizations of
  the $r$-matrix (\ref{cw14bis}) one can use
 the theorem by Belavin and Drinfeld which states that the
  sum  of two $r$-matrices $r_1,r_2$ is again a classical
   $r$-matrix \cite{beldri} if $r_2$ has a carrier
    $L \in g_2$ ($r_2 \in L \otimes L$)
   which cocommutes with $r_1$ (i.e. it is a kernel of the bialgebra cobracket).

   The maximal subalgebra in $g_2$ which is kernel of the Lie bialgebra
   cobracket determined by  the
    $r$-matrix (\ref{cw14bis}) has the following
     linear basis
    \begin{eqnarray}\label{2.xx}
L = (h_{0,1}, e_{0,1}, e_{0,-1}, e_{2,3} )\, .
\end{eqnarray}
i.e. $[r_{2,3;2}(\xi)$, $l\otimes1 +1\otimes l]=0$ ($l\in L$).
From the generators of the
  subalgebra $L$ one can construct the following  five classical
 $r$-matrices:

 a) $h_{0,1}\wedge e_{0,1}$,  \  b) $h_{0,1}\wedge e_{2,3}$, \
  c) $e_{0,1}\wedge e_{2,3}$, \  d) $h_{0,1}\wedge e_{0,-1}$,
\  e) $e_{0,-1}\wedge e_{2,3}$.

The $r$-matrices
 % (\ref{ws2.19}) and (\ref{ws2.21})
which we shall consider below
 are obtained as
 the  linear combination of (\ref{cw14bis}) and the $r$-matrices
  a) and b). One can show that the results of addition of the $r$-matrix
    (\ref{cw14bis})  and
   the $r$-matrices c)--e) can be obtained from the previous two
  cases by suitable automorphisms of the algebra $g_2$.

 It follows that we can consider two $r$-matrices
 as basic ones, or more explicitly:
\bl
 \begin{eqnarray}\label{ws2.24a}
    r_1 & = & \alpha \, h_{0,1} \wedge e_{0,1} + \xi
    \left( h_{2,3} \wedge e_{2,3} +
    e_{1,1} \wedge e_{1,2} + e_{1,3} \wedge e_{1,0}
    \right)  \, ,
\\ \cr
    r_2 & = & \beta \, h_{0,1} \wedge e_{2,3} + \xi
    \left( h_{2,3} \wedge e_{2,3} +
    e_{1,1} \wedge e_{1,2} + e_{1,3} \wedge e_{1,0}
    \right)  \, ,
    \label{ws2.24b}
\end{eqnarray}
\el
where $\xi, \alpha, \beta$  are arbitrary.

One can raise the question  whether  the classical
$r$-matrices (\ref{ws2.24a},b) %--\ref{ws2.24b})
 can be extended to  carrier space containing also the generators  belonging to
 ${\mathfrak{b}}_-$. Unfortunately such an extension,
 which can not be eliminated by the inner automorphism
  of $g_2$, is not possible from purely
 algebraic reason. One can show that there does not exist an even
 dimensional subalgebra of $g_2$,  with dimension ten (two
 extra   generators from ${\mathfrak{b}}_-$), which extends the full Borel
 subalgebra ${\mathfrak{b}}_+$. In fact, the  consideration of
 classical $r$-matrices with the carrier in both Borel
 subalgebras of $g_2$ which however are not simultaneously  the
   classical r-matrices for $sl(3)$
 subalgebra is an interesting problem to study, going beyond
  the scope of the present paper.

 Below  we shall consider the quantization of $g_2$ in the
   four steps, corresponding to the quantization of the
 following sequence of $r$-matrices:

 i) Jordanian twist quantization
\begin{equation}\label{mb2.14bis}
  r_J = \xi \,  h_{2,3}  \wedge  e_{2,3} \, .
\end{equation}

ii) Two extended Jordanian twist quantizations
\bl
\begin{eqnarray}\label{mb2.15abis}
  r_{EJ} &= &\xi \, ( h_{2,3}  \wedge  e_{2,3}  +  e_{1,1}  \wedge  e_{1,2} )\, ,
  \\ \label{mb2.15bbis}
r_{E'EJ} &=& \xi \, ( h_{2,3}  \wedge  e_{2,3}  +  e_{1,1}  \wedge  e_{1,2}  +  e_{1,3}
 \wedge  e_{1,0}   )\, .
\end{eqnarray}
\el

 The $r$-matrix $r_{EJ}$ describes the extended Jordanian twist
quantization of the $sl(3)$  subalgebra.

 iii) Full twist quantization   with additional twist factors
 describing deformed Jordanian
  twist (classical $r$-matrix (\ref{ws2.24a})) and the
  Abelian twist (classical $r$-matrix (\ref{ws2.24b})).

It should be observed that the parameters $\alpha, \beta$ and $\xi$ occurring
 in the classical $r$-matrices (\ref{ws2.24a},b) %--\ref{ws2.24b})
  can be rescaled
 by inner automorphisms of $g_2$ algebra as well as
  by the overall scaling of the $r$-matrices. In particular performing the
 two-parameter rescaling by Cartan generators (we use
  the notation $(a d ^\otimes\,
  a)A \otimes B\equiv [a,A]\otimes B + A \otimes [a, B]$).

\begin{eqnarray}\label{bm2.16}
&&\exp      [a d ^\otimes
(c_1  h_{1,0}  + c_2 h_{0,1} )]
% \cdot\circ
 r_1 = e^{\frac{1}{2}c_1}
%\cdot\circ
 r_1 \, , \cr\cr && \exp [a
 d^\otimes
(c_1  h_{1,0}  + c_2
h_{0,1} )] r_2 =
 \cr\cr
&& =
e^{(-\frac{1}{2}c_1 +\frac{1}{3}c_2)} \,\beta\,
  h_{0,1}  \wedge  e_{0,1}  + e^{\frac{1}{2}c_1}
  %\cdot\circ
   r_{E'EJ}
\end{eqnarray}
we see that while
  the parameter $\alpha$
   remains unchanged,  the parameters $\beta$ and $\xi$
  can be rescaled e.g. to unity.
In order to modify the parameter $\alpha$ we can employ the overall scaling
 of the $r$-matrix.
   We see therefore, that similarly like in the
  case of Jordanian deformation of  $sl(2)$ or $\kappa$-deformation of
  Poincar\'{e} algebra, the deformations with different values of the parameters
   $\alpha, \beta$ and $\xi$ are mathematically
     equivalent (provided $\alpha \neq 0, \beta\neq 0, \xi\neq 0$) but distinguishable
      if applied to physical models.

\section{Twist Quantization Method and the General Twist Functions for
${\bf g}_2$}
\setcounter{equation}{0}

\subsection{Quantum deformations by twisting coproducts of Universal enveloping algebras}

Consider the universal enveloping algebra $U(\mathfrak{g})$ of a Lie algebra
$\mathfrak{g}$ as a Hopf algebra with the comultiplication $\Delta^{(0)}$
generated by the primitive coproduct  in $\mathfrak{g}$. The
parametric invertible solution $\F(\xi) = \sum f^{(1)}_{i} \otimes
f^{(2)}_{i} \in U(\mathfrak{g})\otimes U(\mathfrak{g})$
 of the twist equations \cite{7basic}
\begin{eqnarray}
\label{mb3.1}
 \F_{12}(\Delta^{(0)}\otimes 1)(\F)
=\F_{23}(1\otimes\Delta^{(0)})(\F),
 \\ \cr
\label{mb3.2}
 (\epsilon \otimes {\rm id}) (\F) = ({\rm id} \otimes
\epsilon)(\F) = 1 \otimes 1\, ,
\end{eqnarray}
defines the deformed (twisted) Hopf algebra  $U_{\F}(\mathfrak{g})$ with the
unchanged multiplication, unit and counit (as in  $U(\mathfrak{g})$), the
twisted comultiplication and antipode defined by the relations
\bl
\begin{eqnarray}
\label{mb3.3}
 \Delta_{\F}(u)&=&\F \Delta^{(0)}(u)\F^{-1}\, , \quad u
\in U(\mathfrak{g}), \\ \cr
 S_{\F}(u)&=
  & v \, S^{(0)} (u) \, v^{-1},
 %%%-v u v^{-1},
 \quad
v=\sum f^{(1)}_{i}S^{(0)}(f^{(2)}_{i})\, .
\end{eqnarray}
\el
The twisted algebra  $U_{\F}(\mathfrak{g})$ is triangular, with the
universal ${\cal R}$-matrix
\begin{equation}
\label{mb3.4}
 {\cal R}_{\F}=\F_{21}\F^{-1}\, ,
\end{equation}
which belongs to some extension of $U({\mathfrak{g}})\otimes U({\mathfrak{g}})$.
When $\F$ is a smooth function of $\xi$ and $\lim_{\xi
\rightarrow 0} \F= 1 \otimes 1$ then in the neighborhood of the
origin the ${\cal R}$-matrix can be presented as
\begin{equation}
\label{mb3.5}
 {\cal R}_{\F}= 1 \otimes 1 + \xi r_{\F} + o(\xi)\, ,
\end{equation}
where $ r_{\F}$ is the skewsymmetric classical $r$-matrix
corresponding to the twist $\F$.
Let us write explicitly  the $r$-matrix as follows:
\begin{equation}
\label{mb3.6}
 r_{\cal F} = a^{ij}\, I_i \wedge I_j\, \, .
\end{equation}
Then we obtain
\begin{equation}
\label{mb3.7}
 {\cal F} = 1 \otimes 1 +   \xi \, \tilde{a}^{ij} I_i \otimes
 I_j + {\cal O} (\xi) \, ,
\end{equation}
where $a^{ij} = \frac{1}{2}(\tilde{a}^{ij} - \tilde{a}^{ji})$.

By a nonlinear change of basis in $U(\mathfrak{{g}})$ one can modify the twisted
coproducts and locate  part of the deformation in the algebraic
sector.

\subsection{Twist deformations for $U(g_{2})$ Hopf algebra}

Our aim is to construct explicitly such a sequence of the twist deformations
$U_{{{\cal F}}}(g_{2})$ of the algebra $U(g_{2})$ that will lead to the largest
possible carrier subalgebra for the corresponding classical $r$-matrices.
 The
final element of the corresponding twists will be the full chain of extended
twists
whose carrier coincides with the Borel subalgebra of $g_{2}$.
The peculiarity of the chain twist deformation is that the deformed algebra
can be twisted
 step by step by the consecutive  twisting factors
   with their specific properties.
One of the important aims will be also the construction of proper
nonlinear basis in $U(g_2)$.
Indeed, on  each step we shall construct the nonlinear
basis in which the costructure
of the Hopf algebra $U_{{{\cal F}}}(g_{2})$ becomes more  transparent.

In Sect.~II we have presented the sequence of classical $r$-matrices for $%
U(g_{2})$ (see (\ref{mb2.14bis}), (\ref{mb2.15abis},b) %--\ref{mb2.15bbis})
 and (\ref{ws2.24a},b)). %-\ref{ws2.24b})).
 The quantization of
 these classical $r$-matrices
 is performed as follows.

a) Firstly we introduce the standard Jordanian twist quantizing the
classical $r$-matrix (\ref{mb2.14bis}), corresponding to the
 long root
 $2\alpha_1 +3\alpha_2$
  in $g_2$. We  have the following twisting element \cite{17basic}

%\[
\begin{equation}\label{3.8bis}
{\mbox{${\cal F}$}}_{J}=e^{  h_{2,3}  \otimes \sigma _{2,3}}=e^{H\otimes \sigma }\,,
%\]
\end{equation}
where%
%\[
\begin{equation}
H= h_{2,3}  = 2 h_{1,0} +3 h_{0,1} \,,\qquad \sigma=\ln (1+e_{2,3})\,.
%\]
\end{equation}

b) There are
four types of the extension twisting factors that can be
applied to $%
U_{J}(g_{2})$ \cite{8basic}:%  %%!
\begin{eqnarray}
\label{ab3.10}
{\mbox{${\cal F}$}}_{E_{+}} &=&e^{e_{1,1}\otimes e_{1,2}e^{-\frac{1}{2}\sigma }},
\cr
 {\mbox{${\cal F}$}}_{E_{-}} &=&e^{-e_{1,2}\otimes e_{1,1}e^{-\frac{1}{2}\sigma
}}, \cr
\widetilde{\mbox{${\cal F}$}}_{E_{+}} &=&e^{e_{1,3}\otimes
e_{1,0}e^{-\frac{1}{2}\sigma }}, \cr
\widetilde{\mbox{${\cal F}$}}_{E_{-}} &=&e^{-e_{1,0}\otimes  e_{1,3} e^{-\frac{%
1}{2}\sigma }}\, .
\end{eqnarray}%
They can be composed to provide the following four types of the  %
 two-element extensions of (\ref{3.8bis})
\begin{eqnarray}
\label{ws3.11}
{\mbox{${\cal F}$}}_{E_{++}} &=&e^{e_{1,3}\otimes e_{1,0}e^{-\frac{1}{2}\sigma
}}e^{e_{1,1}\otimes e_{1,2}e^{-\frac{1}{2}\sigma }}, \cr
{\mbox{${\cal F}$}}_{E_{+-}} &=&e^{e_{1,3}\otimes e_{1,0}e^{-\frac{1}{2}\sigma
}}e^{-e_{1,2}\otimes  e_{1,1} e^{-\frac{1}{2}\sigma
}},
\cr
 {\mbox{${\cal F}$}}_{E_{-+}}
 &=
 &e^{- e_{1,0}\otimes e_{1,3}
e^{-\frac{1}{2}\sigma }}e^{e_{1,1}\otimes
e_{1,2}e^{-\frac{1}{2}\sigma }},
\cr
 {\mbox{${\cal F}$}}_{E_{--}}
&=&e^{- e_{1,0}\otimes e_{1,3} e^{-\frac{1}{2}\sigma
}}e^{-e_{1,2}\otimes  e_{1,1} e^{-\frac{1}{2}\sigma
}}.
\end{eqnarray}%
One can note that exponential factors in the twists
(\ref{ws3.11}) commute with each other, and do
not describe themselves the solutions of twist equations
\hbox{(\ref{mb3.1},\ref{mb3.2})} with primitive coproduct $\Delta^{(0)}$.
The four twists (\ref{ws3.11})
 lead to
 the equivalent Hopf algebras
however their coalgebra relations  differ considerably. The most
elegant result is obtained when the extension is chosen as follows
%\[
\begin{equation}
{\mbox{${\cal F}$}}_{E}:=
{\mbox{${\cal F}$}}_{E_{-+}}=
{\mbox{${\cal F}$}}_{E_{-}}
{\mbox{${\cal F}$}}_{E_{+}}=
e^{-e_{1,2}\otimes  e_{1,1} e^{-\frac{1}{2}\sigma }}
e^{e_{1,3}\otimes e_{1,0}e^{-\frac{1}{2}\sigma }}
\, ,
%\]%
\end{equation}
with the extended twist%
%\[
\begin{equation}\label{ws3.13}
{\mbox{${\cal F}$}}_{EJ}:=
e^{-e_{1,2}\otimes e_{1,1}e^{-\frac{1}{2}\sigma }}
e^{e_{1,3}\otimes e_{1,0}e^{-\frac{1}{2}\sigma }}
e^{H\otimes \sigma }
\, .
%\]
\end{equation}
It should be added that the products of twists ${\mbox{${\cal F}$}}_{E_{\pm}}
 {\mbox{${\cal F}$}}_{J}$
describe the twist quantization of $sl(3)$ subalgebra.

c) The additional Abelian twist factor
 ($h\equiv 3  h_{0,1} $)%
%\[
\begin{equation}
{\mbox{${\cal F}$}}_{A}=e^{h\otimes \sigma }\, ,
%\]%
\end{equation}
 that produces a kind of a "rotation" in the root
space of $g_{2}$,
 can enlarge the extended twist (\ref{ws3.13}):
\begin{equation}
\label{ws3.15}
{\mbox{${\cal F}$}}_{AEJ}:=
e^{h\otimes \sigma }
e^{-e_{1,2}\otimes e_{1,1}e^{-\frac{1}{2}\sigma }}\,
e^{e_{1,3}\otimes e_{1,0}e^{-\frac{1}{2}\sigma}}\,
e^{H\otimes \sigma }\, .
%\]
\end{equation}
In such a way we obtain the quantization of the classical
$r$-matrix (\ref{ws2.24b}).

d) We can construct the chain of twists (see e.g.
\cite{KLO,Kul-Lya})
 for $g_{2}$ by additionally
deforming the twisted $U_{EJ}(g_{2})$ by the second link of the chain, which
is the Jordanian factor:%
%\[
\begin{equation}
{\mbox{${\cal F}$}}_{J^{\prime }}=e^{h\otimes \omega }
%\]%
\end{equation}
with%
%\[
\begin{equation}\label{ws3.17}
\omega =\ln \left( 1+e_{0,1}+\frac{1}{2}\left( e_{1,2}\right) ^{2}\right) \, .
%\]%
\end{equation}
This gives the quantization with the largest carrier
%\[
\begin{equation}
\label{ll2.1}
{\mbox{${\cal F}$}}_{J^{\prime }EJ}:=
e^{h\otimes \omega }
e^{-e_{1,2}\otimes e_{1,1}e^{-\frac{1}{2}\sigma }}\,
e^{e_{1,3}\otimes e_{1,0}e^{-\frac{1}{2}\sigma }}\,
e^{H\otimes \sigma }.
%\]
\end{equation}

The twist function (\ref{ll2.1}) describes the quantization of the
classical $r$-matrix (\ref{ws2.24a}).
 The  twist (\ref{ws3.15}) can also form the  chain with
  ${\mbox{${\cal F}$}}_{J''} = e^{h\otimes \omega''}$ (see Section~V.C).
   But the Abelian twist
   factors ${\mbox{${\cal F}$}}_{A}$ and this new  Jordanian factor are
     related by the formula
    ${\mbox{${\cal F}$}}_{J''} {\mbox{${\cal F}$}}_A
    = {\mbox{${\cal F}$}}_{J'}$. This means that for any ``rotated''
    extended twist ${\mbox{${\cal F}$}}_{AEJ}$
     we get the unique chain (\ref{ll2.1}).

In the  Sections~IV and V we shall present the deformed Hopf algebras % male gggggggggg
$U_{{{\cal F}}}(g_{2})$ with more details, discuss their properties and introduce the
suitable bases of $U(g_2)$.
 In such a
way we obtain the  twist quantizations for $g_2$ with the largest
carrier, which can be described by the chain of twists.

The following paragraph (III.C)
will be devoted  to the description of the mathematical framework, which
 permits to choose the basis in $U(g_2)$ with simplified coproduct
  formulae.
  Subsequently, in the paragraph III~D we shall derive some new
mathematical formulae simplifying the calculation of coproducts in
 Sect.~IV and V.
%%% 11-12 Lyak
\subsection{Dual bases and simplification of coalgebra structure}

Let $\left( \mathfrak{g},\mathfrak{g}^{\#}\right) $ be a coboundary
Lie bialgebra. The dual Lie algebra $\mathfrak{g}^{\#}$ is determined
by the $r$-matrix and can be written
explicitly in the form of Lie coalgebra with the cocommutators $\delta \left(
a\right) =\left[ a\otimes 1+1\otimes a,r\right] $, \ ($a\in \mathfrak{g}$).
Let $\frak{G}^{\#}$ denote the dual group for $\left( \mathfrak{g},%
\mathfrak{g}^{\#}\right) $ -- the universal covering Lie group
with the Lie algebra $\frak{g}^{\#}$. According to the quantum
duality principle \cite {9basic} the Hopf algebra
$U_{{{\mathcal{F}}}}(\frak{g})$ naturally treated as quantum
algebra with respect to $U(\frak{g})$ can be also considered as a
quantum group with respect to $\mathrm{Fun}\left(
\frak{G}^{\#}\right) $:
$U_{{{\mathcal{F}}}}(\frak{g})\approx \mathrm{Fun}_{\xi }\left( \frak{G}%
^{\#}\right) $. The coproducts in $U_{{{\mathcal{F}}}}(\frak{g})$
describe the deformed group multiplication law of the dual group
$\frak{G}^{\#}$ . These
multiplications are deformed due to the fact that generators in $\mathrm{Fun}%
_{\xi }\left( \frak{G}^{\#}\right) $ are subject to the relations of
$U(\frak{g})$. The undeformed coproducts for $\mathrm{Fun}\left( \frak{G}%
^{\#}\right) $ can be obtained by constructing the second classical limit $%
\mathrm{Fun}_{\xi }\left( \frak{G}^{\#}\right) \longrightarrow \mathrm{Fun}%
\left( \frak{G}^{\#}\right) $ for the Hopf algebra $U_{{{\mathcal{F}}}}(\frak{g})$
\cite{drin3}. Among other important consequences the quantum duality
prescribes the existence of two preferred bases for the Hopf algebra $U_{\mathcal{F}}(%
\frak{g})$ \cite{19basic}: the natural set of generators for $\frak{g}$
(usually they form the Cartan-Weil basis in $\frak{g}$) and the basis natural for $%
\frak{G}^{\#}$. The latter may be the exponential basis, generated by the
Cartan-Weil basis in $\frak{g}^{\#}$. Evidently, the costructure of $U_{{{%
\mathcal{F}}}}(\frak{g})$ becomes transparent only in terms the group $\frak{%
G}^{\#}$ that is in the dual group basis or (keeping in mind the
exponential map) in $\frak{g}^{\#}$-basis. For algebras of ${\rm
rank}(\frak{g})=1$ one can use the $\frak{g}$-coordinates for the
multiplications as well as for the comultiplications. Starting
with ${\rm rank}(\frak{g})=2$  (see for example
$U_{{{\mathcal{F}}}}(sl(3))$ in \cite{8basic}) some of the coproducts
written in classical $\frak{g}$-basis (natural
$\mathfrak{g}$-coordinates) are complicated and it is difficult to
study their properties. To pass from $\mathfrak{g}$- to
$\mathfrak{g}^{\#}$-basis it is necessary to compare the
generators in $U_{\mathcal{F}}(\mathfrak{g})$ induced by the
undeformed $U(\frak{g})$ with those corresponding to $
\mathrm{Fun}\left( \frak{G}^{\#}\right)$ or to its canonical dual
$\left( \mathrm{Fun}\left( \frak{G}^{\#}\right) \right) ^{\ast
}\approx U(\mathfrak{g}^{\#})$. The general description of the
corresponding algorithm and in particular the transformation of
the
Lie-algebra bases (i.e. the change of coordinates $\mathfrak{g}%
\longleftrightarrow \mathfrak{g}^{\#}$) was presented in
\cite{9basic}. In Section~IV for the obtained quantum algebras
$U_{{\mathcal{F}}}(\mathfrak{g})$ we shall use both
$\mathfrak{g}$- and $\mathfrak{g}^{\#}$-bases  and demonstrate
their role in the description of algebraic and coalgebraic
properties of twist deformations.

%%%
\subsection{%
 New algebraic formulae for the similarity transformation of
tensor products}

Calculations of deformed coproducts rely on successive application of
the following version of the Baker-Campbell-Hausdorff (BCH) formula
(the adjoint action in terms of exponential map):
\begin{equation}
\label{a3.20}
e^{X}Ae^{-X}=\exp \left( \mathrm{ad}_{X}\right) A\equiv \sum_{k=0}^{\infty }%
\frac{1}{k!}\mathrm{ad}_{X}^{k}A=\sum_{k=0}^{\infty }\frac{1}{k!}%
[X,[..[X,A]..]
\, .
\end{equation}
More exactly, since one works in the tensor product of two algebras $%
U\otimes U$, the BCH formula we need is as follows,
\begin{eqnarray}
\label{a3.21}
e^{X\otimes Y}(A\otimes B)e^{-X\otimes Y} & = &\exp \left( \mathrm{ad}_{X\otimes
Y}\right) (A\otimes B)
\cr\cr
 &=&  % \notag \\
\exp \left( \mathrm{ad}_{X}\otimes Y\right) (A\otimes 1)\exp \left( X\otimes
\mathrm{ad}_{Y}\right) \left( 1\otimes B\right)
\cr\cr
&=& %  \notag \\
\sum_{k=0}^{\infty }\frac{1}{k!}\mathrm{ad}_{X}^{k}A\otimes
Y^{k}\sum_{m=0}^{\infty }\frac{1}{m!}X^{m}\otimes \mathrm{ad}_{Y}^{m}B
\, .
\cr &&
\end{eqnarray}%
The above expression is a product of two infinite series. Fortunately, in
most of applications important for our study here
both series truncate and become finite due to the fact that one
acts in the enveloping algebra \ $U=U_{g}$\ of some
  finite dimensional
simple Lie algebra $g $ and  adjoint actions of non-Cartan elements
are nilpotent.
 The other case which can be handled well occurs
 when one of the elements, say $X$ , belongs to Cartan subalgebra. Then $%
\mathrm{ad}_{X}A\equiv \lbrack X,A]=aA$ \ and $\mathrm{ad}_{X}^{k}A=a^{k}A$
with $a\in
\mathbb{C}
$. In this case the first factor shrinks into
the following simple expression

\begin{equation}
\label{a3.22}
\sum_{k=0}^{\infty }\frac{1}{k!}\mathrm{ad}_{X}^{k}A\otimes Y^{k}=A\otimes
e^{aY}\, .
\end{equation}%
We see that the complexity of calculations heavily depends on
the degree of nilpotency.
 We should also notice that in applications
 one of
the factors, say $Y=f\left( E\right) $ might have
 a  functional form in terms of some
generator  \ $E$, e.g. \ corresponding
 to maximal root vector. We are particularly interested
when $f\left( E\right) =\ln \left( 1+E\right) $\ \ \ or \ \ $f\left(
E\right) =e^{E}$.

When degree of nilpotency is lower then two, i.e. $\mathrm{ad}%
_{E}^{2}B\equiv \lbrack E,[E,B]]=0$
 one can use
 the following
obvious expression $\ [f(E),B]=[E,B]f^{\prime }(E)$ (where $f^{\prime }$\
denotes derivative of $f$) . This leads to
\begin{equation}
\label{a3.23}
\sum_{m=0}^{\infty }\frac{1}{m!}X^{m}\otimes \mathrm{ad}_{f\left( E\right)
}^{m}B=1\otimes B+X\otimes \lbrack E,B]f^{\prime }(E)
\, .
\end{equation}

Dealing with these simple technique permits to calculate almost all
deformed coproducts except the
 one described by the last twist factor ${\cal F}_C$ (see (\ref{blm5.22})).
 In order to complete
calculations one needs more general and more sophisticated methods.
 For this purpose
we have found two combinatorial expressions which turned out to be very
useful.

If $X$ and $Y$ are two commuting elements and in addition $[X,B]=0$ then
\begin{eqnarray}
\label{a3.24}
Y^{X}BY^{-X}\equiv e^{X\ln Y}Be^{-X\ln Y}\ =\sum_{k=0}^{\infty }X^{%
\underline{k}}\frac{\mathrm{ad}_{Y}^{k}B}{k!}Y^{-k}
\end{eqnarray}
where the sequence
\begin{equation}
\label{a3.25}
x^{\underline{k}}\equiv
\frac{\Gamma(x+1)}{\Gamma(x-k+1)} =
x\left( x-1\right) \ldots \left( x-k+1\right)
=\sum_{m=1}^{k}s\left( k,m\right) x^{m}
\end{equation}%
stands for the so-called lower (or falling) factorial polynomials and $%
s\left( n,k\right) $ are the Stirling numbers of the first kind.

Notice that  if $[Y,B]=0$ and \ $[Y,X]=0$ one gets
\begin{equation}
\label{a3.26}
Y^{X}BY^{-X}\equiv e^{X\ln Y}Be^{-X\ln Y}=\sum_{k=0}^{\infty }\frac{1}{k!}%
\left( \ln Y\right) ^{k}{\mathrm{ad}}_{X}^{k}B
\end{equation}%
Another useful combinatorial formula is
\begin{eqnarray}
\label{a3.27}
e^{Ye^{E}}Be^{-Ye^{E}}=\sum_{k=0}\ \frac{\mathrm{ad}_{E}^{k}B}{k!}%
q_{k}\left( Ye^{E}\right)
 =\sum_{m=0}(\sum_{k=m}\frac{\mathrm{ad}_{E}^{k}B}{%
k!}S(k,m))Y^{m}e^{mE}
\end{eqnarray}%
whenever $E$  and  $B$ commute with $Y$.
  Here the sequence
\begin{equation}
\label{a3.28}
q_{k}\left( x\right) =\sum_{m=0}^{k}S(k,m)x^{m}
\end{equation}%
is given by the so-called Bell polynomials and \ $S\left(
n,k\right) $ are Stirling numbers of the second kind. It should be also
remarked that both families of polynomials,
 $x^{\underline{k}}$ and
  $q_k(x)$,  belong to the wide class of
so-called convolution polynomials known also as polynomial sequences of
binomial type since, e.g.
\begin{equation}
\label{a3.29}
q_{k}\left( x+y\right) =\sum_{m=0}^{k}\binom{k}{m}q_{m}\left( x\right)
q_{k-m}\left( y\right)
\end{equation}
Such polynomials are extensively discussed in combinatorial
analysis and umbral calculus (see
\cite{111basic},\cite{222basic}).
 Here we have found operator analogs of some fundamental formulae involving
  polynomial sequences of binomial type.
%%%
 All above formulae can be checked by direct calculation in
any order of the power expansion, the non-perturbative proof will be
presented elsewhere.

Both formulae (\ref{a3.24}) and (\ref{a3.27}) can be adjusted to the
particular situation for non-standard (Jordanian) and \ extended twists.
     For example, adapting to the case of Jordanian twist ${{{\cal F}}}_{J}=e^{H\otimes
\omega }$ with $H$\ \ being a Cartan element: $[H,A]=aA$, $\omega =\ln
\left( 1+E\right) $ , \ one gets
\begin{equation}
\label{a3.30}
{{{\cal F}}}_{J}A\otimes B{{{\cal F}}}_{J}^{-1}
=\ \sum_{k=0}^{\infty }H^{\underline{k}}\otimes
\frac{\mathrm{ad}_{E}^{k}B}{k!}e^{-k\omega }\left( A\otimes e^{a\omega
}\right) =\sum_{k=0}^{\infty }H^{\underline{k}}A\otimes \frac{\mathrm{ad}%
_{E}^{k}B}{k!}e^{(a-k)\omega }
\, .
\end{equation}

Similarly, the counterpart of (\ref{a3.30})
 for extended twist (double exponential)\ $%
{{{\cal F}}}_{E}=e^{X\otimes Ye^{f(E)}}$\ takes the form
\begin{eqnarray}
\label{a3.31}
{{{\cal F}}}_{E}(A\otimes B){{{\cal F}}}_{E}^{-1}
\!\!\!\!\!\!
&=&
\!\!\!\!\!
 \sum_{m=0}^{\infty }\frac{1}{m!}%
X^{m}\otimes \mathrm{ad}_{f(E)}^{m}B\ q_{m}\left( Ye^{f(E)}\right)
\sum_{k=0}^{\infty }\frac{1}{k!}\mathrm{ad}_{X}^{k}A\otimes Y^{k}e^{k\, f(E)}
\notag \\
&=&
\!\!\!\!\!
\sum\limits_{k,m\geq 0}\frac{1}{m!k!}X^{m}\mathrm{ad}_{X}^{k}A\otimes
\mathrm{ad}_{f(E)}^{m}B\sum_{i=0}^{m}S(m,i)Y^{k+i}e^{\left( k+i\right) f(E)}
\cr &&
\end{eqnarray}%
provided that \ $[Y,B]=[Y,E]=0$ .

In Sections~IV and V these last formulae shall be
particularly useful.

\section{Twist Quantizations of $g_{2}$ Generated by the $sl(3)$ Twists.}
\setcounter{equation}{0}

We start by considering the two twist factors ${\mbox{${\cal F}$}}$
 corresponding to the classical $r$-matrix (\ref{mb2.15abis}) with
the carrier subalgebras
 inside the Borel subalgebra of  $sl(3)\subset g_2$.
%  we recall that  $sl(3)\subset g_{2}$.
  This choice will permit to
 construct first two steps of the twist deformation
  corresponding to the $r$-matrices (\ref{ws2.24a},b). %--\ref{ws2.24b}).
Performing such twisting  one can
compare deformed Hopf algebras $U_{{{{\cal F}}}}(g_{2})$  with the
 known deformation $U_{{{{\cal F}}}}(sl(3))$ \cite{7basic,8basic}.

\subsection{The First Jordanian Twist}
The Jordanian twist ${{{\cal F}}}_{J}$ can be based on any 2-dimensional
Borel subalgebra ${\mathfrak{b}}_{+}(sl(2))\subset g_{2}$.
 In our case this subalgebra is  generated by
 $\left\{ H\equiv h_{2,3},e_{2,3}\right\}$.
  The Jordanian $r$-matrix is given by (\ref{mb2.14bis}), and
  the corresponding twisting element is
$ %\begin{equation}
{{{\cal F}}}_{J} %=e^{\left( 2h_{1}+3h_{2}\right) \otimes \sigma_{2,3}}
=e^{H\otimes \sigma }\,
$%\end{equation}
(see (\ref{3.8bis}))

The costructure of the deformed algebra $U_{J}(g_{2})$ is
obtained by applying the similarity map (\ref{mb3.3}) to the
primitive coproducts  in $U(g_{2})$. In the
carrier subalgebra
${\mathfrak{b}}_+(sl(2))$ %(Borel subalgebra of $Sl(2)$)
we obtain

\begin{eqnarray}
\label{4.2}
\Delta _{J}\left( e_{2,3}\right)  &=&e_{2,3}\otimes e^{\sigma
}+1\otimes e_{2,3},\quad \mathrm{or\quad }\Delta _{J}\left(
\sigma \right) =\sigma
\otimes 1+1\otimes \sigma ,
\\  \cr
\Delta _{J}\left( H\right)  &=&H\otimes e^{-\sigma }+1\otimes H\, .
\label{4.3}
\end{eqnarray}
The generators of the long root sequences $\pm \left( \alpha_1 + k
\alpha_2 \right)$ $\left( k=0,1,2,3 \right) $ have the coproducts
\begin{equation}
\label{4.4}
\Delta _{J}\left( e_{1,k}\right) =e_{1,k}\otimes
e^{\frac{1}{2}\sigma }+1\otimes e_{1,k}\, ,
%\]%
\end{equation}
\begin{equation}
\label{4.7}
\Delta _{J}\left( e_{-1,-k}\right) =e_{-1,-k}\otimes
e^{-\frac{1}{2}\sigma }+1\otimes e_{-1,-k}+\left( -1\right)
^{k}H\otimes e_{1,\left( 3-k\right) }e^{-\sigma }\, ,
%\]%
\end{equation}
describing 1- and 2- dimensional subrepresentations of the Borel subalgebra
${%\widehat
{\mathfrak{b}}}_+(sl(2))$. The generators with the roots "orthogonal" to
$2\alpha_1 +3\alpha_2 $ remain primitive:
\begin{eqnarray}
\label{4.5}
\Delta _{J}\left( h \right)  &=&h \otimes 1+1\otimes h \, , \\
\label{4.6}
\cr
\Delta _{J}\left( e_{0,\pm 1}\right)  &=&e_{0,\pm 1}\otimes
1+1\otimes e_{0,\pm 1}\, .
\end{eqnarray}%
and describe 2-dimensional subrepresentations of ${{{%\widehat
{\mathfrak{b}}}}}_+(sl(2))$.
One can check that in ${\mathfrak{g}}^{\#}\setminus
 {\mathfrak{b}}_+(sl(2))$ the generator $H$ induces a
shift
 $\mathrm{ad}_{H}:e_{1,\left( 3-k\right) }\longrightarrow
e_{-1,-k}$ and this is exactly what indicates the last terms of the above
coproducts $\Delta _{J}\left( e_{-1,-k}\right)$.

The coproduct for the lowest root generator $e_{-2,-3}$,%
\begin{eqnarray}
\label{3.8}
 \Delta _{J}\left( e_{-2,-3}\right) &=&e_{-2,-3}\otimes
e^{-\sigma }+1\otimes e_{-2,-3}
\cr\cr
 &&+\left( H-H^{2}\right)
\otimes \left( e^{-\sigma }-e^{-2\sigma }\right) +2H\otimes
He^{-\sigma }\, ,
\end{eqnarray}%
also refers to a 1-dimensional subrepresentation. This can be seen
when we pass to the  $\mathfrak{g}^{\#}$ -basis.
The following generator should be redefined here,

\begin{equation}
\widetilde{e_{-2,-3}}:=e_{-2,-3}-H^{2}\, .  \label{new-c-1}
\end{equation}%
The coproduct $\Delta _{J}\left( \widetilde{e_{-2,-3}}\right) $
is quasiprimitive and similar to $\Delta_{ J}(H)$:
\begin{equation}
\Delta _{J}\left( \widetilde{e_{-2,-3}}\right) =
\widetilde{e_{-2,-3}}%
\otimes e^{-\sigma }+1\otimes \widetilde{e_{-2,-3}}\, .
\end{equation}
For the case of twisted $U_{J}\left( sl(2)\right) $ the nonlinear
transformation (\ref{new-c-1}) was first indicated in
\cite{17basic}.

\subsection{The First Extended Jordanian Twist}

The peculiarity of the chain twist deformation is that the
deformed algebra can be twisted further by other twisting
factors. The quantization goes step by step and on each level you
get the deformed symmetry with its specific properties.

For the carrier $\mathfrak{g}^{c}\subset sl(3)\subset g_{2}$
 the extension factor should be
\begin{equation}
\label{co-e+-0}
 {\mbox{${\cal F}$}}_{E_{+}}=e^{e_{1,3}\otimes
e_{1,0}e^{-\frac{1}{2}\sigma }}\, .
\end{equation}

We remind that $sl(3)$ subalgebra is generated by the
following  8 elements
$\{ H, h, e_{\pm 2,\pm 3}, e_{\pm 1, 0}, e_{\pm 1, \pm 3}\}$.

The twist (\ref{co-e+-0}) is a solution of the twist equations
 (\ref{mb3.1}), (\ref{mb3.2}) for
$U_{J}(g_{2})$ and the adjoint operator $\exp \left(
\mathrm{ad}\left( e_{1,3}\otimes e_{1,0}e^{-\frac{1}{2}\sigma
}\right) \right) $ applied to the coproducts in
$U_{J}(g_{2})$ will perform the deformation
${\mbox{${\cal F}$}}_{E_{+}}:U_{J}(g_{2})%
\longrightarrow U_{E_{+}J}(g_{2})$.  The same result
can be
obtained directly by applying the twist
${\mbox{${\cal F}$}}_{E_{+}J}=\left(
{\mbox{${\cal F}$}}_{E_{+}}{\mbox{${\cal F}$}}_{J}\right):
 U(g_{2})\longrightarrow U_{E_{+}J}(g_{2})$.

In the costructure of $U_{E_{+}J}(g_{2})$ we see the group
multiplication of the
solvable 4-dimensional Lie group
(see \cite{8basic})
 with the Lie algebra
equivalent to the carrier subalgebra $\mathfrak{g}_{E_{+}J}^{c}$ of the twist
$%
{\mbox{${\cal F}$}}_{E_{+}J}$:%

\begin{eqnarray}
\label{co-e+-01}
\Delta _{E_{+}J}\left( e_{2,3}\right)  &=&e_{2,3}\otimes e^{\sigma
}+1\otimes e_{2,3},   \cr
\Delta _{E_{+}J}\left( H\right)  &=&H\otimes e^{-\sigma }+1\otimes
H-e_{1,3}\otimes e_{1,0}e^{-\frac{3}{2}\sigma },
\nonumber \\
\Delta _{E_{+}J}\left( e_{1,0}\right)  &=&e_{1,0}\otimes e^{\frac{1}{2}%
\sigma }+e^{\sigma }\otimes e_{1,0},
 \nonumber \\
\Delta _{E_{+}J}\left( e_{1,3}\right)  &=&e_{1,3}\otimes e^{-\frac{1}{2}%
\sigma }+1\otimes e_{1,3}\, .
%\label{co-e+-01}
\end{eqnarray}

The other generators of the subalgebra $sl(3)$ form the 4-dimensional
representation of $\mathcal{L}$:
\begin{eqnarray}
\Delta _{E_{+}J}\left( h\right) &=&h \otimes 1+1\otimes h \, ,
\label{co-e+-02}
\\
\Delta _{E_{+}J}\left( e_{-1,-3}\right) &=&e_{-1,-3}\otimes e^{-\frac{1}{2}%
\sigma }+1\otimes e_{-1,-3}+ h\otimes e_{1,0}e^{-\sigma }\, ,
\nonumber
\\
\Delta _{E_{+}J}\left( e_{-1,0}\right) &=&+e_{-1,0}\otimes e^{-\frac{1}{2}%
\sigma }+1\otimes e_{-1,0}+H\otimes e_{1,3}e^{-\sigma }
 \nonumber
 \\
&&-e_{1,3}\otimes e_{1,0}e_{1,3}e^{-\frac{3}{2}\sigma
}+e_{1,3}\otimes \left( H-h\right) e^{-\frac{1}{2}\sigma }
\nonumber
\\
&&-\left( e_{1,3}\right) ^{2}\otimes e_{1,0}e^{-2\sigma
}-He_{1,3}\otimes
\left( e^{\sigma }-1\right) e^{-\frac{3}{2}\sigma }\, ,
 \nonumber
  \\
\Delta _{E_{+}J}\left( \widetilde{e_{-2,-3}}\right) &=&+\widetilde{e_{-2,-3}}%
\otimes e^{-\sigma }+1\otimes
\widetilde{e_{-2,-3}}-e_{-1,0}\otimes
e_{1,0}e^{-\frac{3}{2}\sigma }
 \nonumber \\
&&+He_{1,3}\otimes e_{1,0}e^{-\frac{3}{2}\sigma }+e_{1,3}\otimes
e_{-1,-3}e^{-\frac{1}{2}\sigma }\, .  \nonumber
\end{eqnarray}

Finally on the remaining 6-dimensional space we also observe the adjoint
action of the carrier subalgebra ${\mathfrak{g}}^{c}_{E_+J}$ ($k=1,2$)
\begin{eqnarray}
\Delta _{E_{+}J}\left( e_{1,k}\right)  &=&e_{1,k}\otimes e^{\frac{1}{2}%
\sigma }+1\otimes e_{1,k}\, ,   \nonumber \\
\Delta _{E_{+}J}\left( e_{0,\pm 1}\right)  &=&e_{0,\pm 1}\otimes
1+1\otimes
e_{0,\pm 1}+e_{1,\frac{5\pm 1}{2}}\otimes e_{1,\frac{1\pm 1}{2}}e^{-\frac{1}{%
2}\sigma }\, ,  \nonumber \\
\Delta _{E_{+}J}\left( e_{-1,-k}\right)  &=&e_{-1,-k}\otimes e^{-\frac{1}{2}%
\sigma }+1\otimes e_{-1,-k}+\left( -1\right) ^{k}H\otimes
e_{1,\left(
3-k\right) }e^{-\sigma }  \nonumber \\
&&-e_{\left( 2-k\right) ,\left( 5-2k\right) }\otimes e_{\left(
k-1\right) ,\left( k-2\right) }e^{-\sigma }
\nonumber \\
&&
- \ e_{1,3}\otimes
e_{1,0}e_{1,\left( 3-k\right) }e^{-\frac{3}{2}\sigma }\, .
% \nonumber
\label{co-e+-03}
\end{eqnarray}

First of all notice that $sl(3)$  generates the well
known twisted algebra $U_{EJ}(sl(3))$ \sloppy
 \cite{7basic}
  which here is a Hopf subalgebra
$U_{EJ}(sl(3))\subset U_{E_{+}J}(g_{2})$.
 Thus we obtain the
intermediate $U_{{{\cal F}}}\left( sl(3)\right) $ twist quantization
inside $U_{{{\cal F}}}(f_{2})$.

The nonprimitive terms in $\Delta _{E_{+}J}\left( e_{0,\pm
1}\right)$ are in agreement with the structure of $\mathfrak{g}^{\#}$. They
describe the action of the dual carrier group $G_{c}^{\#}$ in the
2-dimensional indecomposable
representations.
The third term in $%
\Delta _{E_{+}J}\left( H\right) $ (see (\ref{co-e+-01})) is just due to the
Heisenberg subgroup in $G_{c}^{\ast }$.

The essential nonlinearities in the costructure are present in
  the last terms in $%
\Delta _{E_{+}J}\left( e_{-1,-k}\right) $
 as well as in $\Delta _{E_{+}J}
\left( \widetilde{e_{-2,-3}}\right) $ and  $%
\Delta _{E_{+}J}\left( e_{-1,0}\right) $.
 It should be noticed that the generator
$\widetilde{e_{-2,-3}}= e_{-2,-3} - H^2$ is not modified.
  This is a common property of all
the generators with roots opposite to the Jordanian carrier.

Let us return to the nontrivial terms in the costructure $\Delta
_{E_{+}J}$. Comparing $\Delta _{E_{+}J}\left( e_{-1,0}\right) $,
$\Delta _{E_{+}J}\left( e_{-1,-k}\right) $ and $\Delta
_{E_{+}J}\left( e_{0,\pm 1}\right) $ with the canonical
multiplication in $U\left( \mathfrak{g}^{\#}\right) $ we find
 the following $\mathfrak{g}^{\#}$ -basis:

\begin{eqnarray}
\widetilde{e_{-1,0}} &=&e_{-1,0}-He_{1,3},  \label{tilda-coor} \\
\widetilde{e_{0,-1}} &=&e_{0,-1}-e_{1,0}e_{1,2}e^{-\sigma }\, ,  \nonumber \\
\widetilde{e_{0,1}} &=&e_{0,1}-e_{1,3}e_{1,1}\, ,  \nonumber \\
\widetilde{e_{-1,-2}} &=&e_{-1,-2}-He_{1,1}\, .  \nonumber
\end{eqnarray}%
In these terms
the action of $G^{c}$ on the 6-dimensional space becomes
transparent:

\begin{eqnarray}
\label{4.16}
\Delta _{E_{+}J}\left( \widetilde{e_{0,1}}\right) &=&\widetilde{e_{0,1}}%
\otimes 1+1\otimes \widetilde{e_{0,1}}-e_{1,1}\otimes e_{1,3}e^{\frac{1}{2}%
\sigma }\, ,
\cr
\label{4.17}
\Delta _{E_{+}J}\left( \widetilde{e_{0,-1}}\right) &=&\widetilde{e_{0,-1}}%
\otimes 1+1\otimes \widetilde{e_{0,-1}}-e_{1,0}e^{-\sigma }\otimes
e_{1,2}e^{-\frac{1}{2}\sigma }\, ,
 \cr
 \label{4.18}
\Delta _{E_{+}J}\left( e_{-1,-1}\right) &=&e_{-1,-1}\otimes e^{-\frac{1}{2}%
\sigma }+1\otimes e_{-1,-1}-H\otimes e_{1,2}e^{-\sigma }
\cr
%\label{4.19}
&&-e_{1,3}\otimes \widetilde{e_{0,-1}}e^{-\frac{1}{2}\sigma }\, ,
\cr
\label{4.20}
\Delta _{E_{+}J}\left( \widetilde{e_{-1,-2}}\right) &=&\widetilde{e_{-1,-2}}%
\otimes e^{-\frac{1}{2}\sigma }+1\otimes \widetilde{e_{-1,-2}}%
-e_{1,1}\otimes He^{\frac{1}{2}\sigma }
\cr
%\label{4.21}
&&-\widetilde{e_{0,1}}\otimes e_{1,0}e^{-\sigma }\, ,
\cr
\label{4.22}
\Delta _{E_{+}J}\left( \widetilde{e_{-1,0}}\right) &=&\widetilde{e_{-1,0}}%
\otimes e^{-\frac{1}{2}\sigma }+1\otimes
\widetilde{e_{-1,0}}-e_{1,3}\otimes  he^{-\frac{1}{2}\sigma }\, ,
\cr
\label{4.23}
\Delta _{E_{+}J}\left( \widetilde{e_{-2,-3}}\right) &=&\widetilde{e_{-2,-3}}%
\otimes e^{-\sigma }+1\otimes \widetilde{e_{-2,-3}}
\cr
&&-\widetilde{e_{-1,0}}\otimes e_{1,0}e^{-\frac{3}{2}\sigma
}+e_{1,3}\otimes e_{-1,-3}e^{-\frac{1}{2}\sigma }\, .
\end{eqnarray}

\section{Twist Deformations Specific to $g_{2}$}
\subsection{The Full Extended Twist}

\setcounter{equation}{0}

The Jordanian twist ${{{\cal F}}}_{J}$ (4.1) can be enlarged by
 the second
extension factor ${{{\cal F}}}_{E_{-}}$. Such an extension
  is the special property of $g_{2}$ root
system.
It does not exist for any other rank 2 simple Lie algebra. The following
element is the solution of the twist equations for the Hopf algebra $%
U_{E_{+}J}(g_{2})$,%
\begin{equation}
{{{\cal F}}}_{E_{-}}=e^{-e_{1,2}\otimes e_{1,1}e^{-\frac{1}{2}\sigma }}.
\label{sec-ext}
\end{equation}%
Together with the previously studied twist ${{{\cal F}}}_{E_{+}J}$
 we obtain  the full
 extended twist%
\begin{equation}
{{{\cal F}}}_{EJ}=e^{-e_{1,2}\otimes e_{1,1}e^{-\frac{1}{2}\sigma }}e^{e_{1,3}\otimes
e_{1,0}e^{-\frac{1}{2}\sigma }}e^{H\otimes \sigma }.  \label{full-ext-tw}
\end{equation}%
The carrier subalgebra $\mathfrak{g}_{EJ}^{c}\approx %
\mathfrak{g}_{EJ}^{c\#}$
 is 6-dimensional, it
 contains two Heisenberg subalgebras with  common central element $e_{2,3}$
and the Cartan generator $H$. Applying the twist (\ref{full-ext-tw}) to $%
U(g_{2})$ or the second extension (\ref{sec-ext}) to the Hopf algebra $%
U_{E_{+}J}(g_{2})$ (constructed in the previous Section) we obtain the new
deformed costructure $\Delta _{E_{-+}J}$.
 The coproducts $\Delta _{EJ}:=\Delta _{E_{-+}J}$
 for the generators of $\mathfrak{g}_{EJ}^{c}$
  describe the group
multiplication in $G_{EJ}^{c\#}$ which is defined by the following relations:

\begin{eqnarray}
\Delta _{EJ}\left( e_{2,3}\right)  &=&e_{2,3}\otimes e^{\sigma }+1\otimes
e_{2,3},  \label{co-car-e1} \\
\Delta _{EJ}\left( H\right)  &=&H\otimes e^{-\sigma }+1\otimes
H-e_{1,3}\otimes e_{1,0}e^{-\frac{3}{2}\sigma }+e_{1,2}\otimes e_{1,1}e^{-%
\frac{3}{2}\sigma }.  \nonumber
\end{eqnarray}%
\begin{eqnarray}
\Delta _{EJ}\left( e_{1,l}\right)  &=&e_{1,l}\otimes e^{\frac{1}{2}\sigma
}+e^{\sigma }\otimes e_{1,l}\, ,\quad l=0,1  \label{co-car-e2} \\
\Delta _{EJ}\left( e_{1,2+l}\right)  &=&e_{1,2+l}\otimes e^{-\frac{1}{2}%
\sigma }+1\otimes e_{1,2+l,}  \nonumber
\end{eqnarray}%
These six coproducts are typical for the extended Jordanian twists
 \cite{7basic}:
 each extension adds a summand in $\Delta _{E_{-+}J}\left( H\right) $,
the constituent roots generators remain quasiprimitive
 and the generator $\sigma$           %%%% (see (IV.25) is not
deformed by the extension factors.

On the plane "orthogonal" to the initial root $\lambda _{0}=2\alpha +3\beta $
we find only one primitive generator%
\begin{equation}
\Delta _{EJ}\left( h\right) =h\otimes 1+1\otimes h.  \label{new-copr1}
\end{equation}%
Other two  coproducts are deformed:%
\begin{eqnarray}
\Delta _{E_{-+}J}\left( e_{0,1}\right)  &=&e_{0,1}\otimes 1+1\otimes e_{0,1}
\label{new-copr2} \\
&&-e_{1,2}\otimes e_{1,2}e^{-\frac{1}{2}\sigma }+\frac{1}{2}\left(
e_{1,2}\right) ^{2}\otimes \left( 1-e^{-\sigma }\right)\, , \nonumber \\
\Delta _{E_{-+}J}\left( e_{0,-1}\right)  &=&e_{0,-1}\otimes 1+1\otimes
e_{0,-1}  \nonumber \\
&&-\frac{4}{3}e_{1,1}\otimes e_{1,1}e^{-\frac{1}{2}\sigma }-\frac{2}{3}%
\left( e^{\sigma }-1\right) \otimes \left( e_{1,1}\right) ^{2}e^{-\sigma }.
\nonumber
\end{eqnarray}

The generators in the negative sector
 have quite complicated coproducts in $U_{EJ}(g_{2})$.
For example

\begin{eqnarray}
\Delta _{EJ}\left( \widetilde{e_{-2,-3}}\right)  &=&\widetilde{e_{-2,-3}}%
\otimes e^{-\sigma }+1\otimes \widetilde{e_{-2,-3}}  \label{new-copr7} \\
&&-\left( e_{-1,0}-He_{1,3}\right) \otimes e_{1,0}e^{-\frac{3}{2}\sigma
}+e_{1,3}\otimes e_{-1,-3}e^{-\frac{1}{2}\sigma }
%\nonumber \\
\cr
&&-\left( e_{-1,-1}-\frac{1}{4}e_{1,2}+He_{1,2}\right) \otimes e_{1,1}e^{-%
\frac{3}{2}\sigma }
%\nonumber \\
\cr
&&-\left( \frac{2}{3}e_{0,1}+\frac{1}{4}\left( e_{1,2}\right) ^{2}\right)
\otimes \left( e_{1,1}\right) ^{2}e^{-2\sigma }
% \nonumber \\
\cr
&&+e_{1,2}\otimes \left( e_{-1,-2}-\frac{1}{4}e_{1,1}e^{-\sigma }\right) e^{-%
\frac{1}{2}\sigma }
% \nonumber \\
\cr
&&+\frac{1}{2}\left( e_{1,2}\right) ^{2}\otimes \left( \frac{1}{2}\left(
e_{1,1}\right) ^{2}e^{-\sigma }+e_{0,-1}\right) e^{-\sigma }
%\nonumber \\
\cr
&&+\frac{2}{9}e_{1,3}\otimes \left( e_{1,1}\right) ^{3}e^{-\frac{5}{2}\sigma
}-\frac{1}{6}\left( e_{1,2}\right) ^{3}\otimes e_{1,0}e^{-\frac{3}{2}\sigma
}.  \nonumber
\end{eqnarray}
Notice that in this expression we use the generators
$\widetilde{e_{-2,-3}}$, i. e. we suppose that
  $\widetilde{e_{-2,-3}}$ will be
appropriate for the deformed costructure not only in $U_{E_{+}J}(g_{2})$ but
also in $U_{EJ}(g_{2})$.

The coproducts for the elements $\left\{ h,e_{0,\pm
1},e_{-1,.-k},e_{-2,-3};k=0,\ldots , 3\right\}$  describe the action
of the carrier group $G_{EJ}^{c\#}$ in the 8-dimensional subrepresentation.

%%% Lyak AAAAAAAAAAAAAAAAAA
To make this adjoint action transparent we perform the coordinate
transformation $\mathfrak{g} \Longrightarrow \mathfrak{g}^{\#}$
(see Section III.C). According to the algorithm presented in
\cite{8basic} the new basic elements are introduced:
%%%

\begin{eqnarray}
\widetilde{\widetilde{e_{0,1}}} &=&e_{0,1}+\frac{1}{2}\left( e_{1,2}\right)
^{2}\, ,  \label{new-coord-3}
\\
\widetilde{\widetilde{e_{0,-1}}} &=&e_{0,-1}+\frac{2}{3}\left(
e_{1,1}\right) ^{2}e^{-\sigma }\, ,
  \nonumber \\
\widetilde{\widetilde{e_{-1,-3}}} &=&e_{-1,-3}+\frac{2}{9}\left(
e_{1,1}\right) ^{3}e^{-2\sigma }\, ,  \nonumber
\\
\widetilde{\widetilde{e_{-1,-2}}} &=&e_{-1,-2}\, ,
  \nonumber \\
\widetilde{\widetilde{e_{-1,-1}}} &=&e_{-1,-1}+He_{1,2}\, ,
  \nonumber \\
\widetilde{\widetilde{e_{-1,0}}} &=&e_{-1,0}-He_{1,3}+\frac{1}{6}\left(
e_{1,2}\right) ^{3}\, ,
  \nonumber \\
\widetilde{\widetilde{e_{-2,-3}}} &=&\widetilde{e_{-2,-3}}\, .  \nonumber
\end{eqnarray}%
 We see that
the element $e_{-1,-2}$ belongs to the $\mathfrak{g}^{\#}$-basis for the group
$G_{EJ}^{\#}$ while
  the generator $\widetilde{e_{-2,-3}}$ remains unchanged  however
   we recall that
 on the previous step of
quantization (in $U_{E_{+}J}\left( g_{2}\right) $) it was nontrivially
deformed.    The coproducts of generators (\ref{new-coord-3})
 are the following
\begin{eqnarray}
\Delta _{EJ}\left( \widetilde{\widetilde{e_{0,1}}}\right)  &=&\widetilde{%
\widetilde{e_{0,1}}}\otimes 1+1\otimes \widetilde{\widetilde{e_{0,1}}}\, ,
\label{new-copr-2}
\cr
\Delta _{EJ}\left( \widetilde{\widetilde{e_{0,-1}}}\right)  &=&\widetilde{%
\widetilde{e_{0,-1}}}\otimes 1+1\otimes \widetilde{\widetilde{e_{0,-1}}}\, ,
\cr
\Delta _{EJ}\left( h\right)  &=&h\otimes 1+1\otimes h
\cr
\Delta _{EJ}\left( \widetilde{\widetilde{e_{-1,-3}}}\right)  &=&\left(
\widetilde{\widetilde{e_{-1,-3}}}\right) \otimes e^{-\frac{1}{2}\sigma
}+1\otimes \left( \widetilde{\widetilde{e_{-1,-3}}}\right)
 \cr
&&+3h_{0,1}\otimes e_{1,0}e^{-\sigma }+\widetilde{\widetilde{e_{0,-1}}}%
\otimes e_{1,1}e^{-\sigma }\, ,
\cr
\Delta _{EJ}\left( e_{-1,-2}\right)  &=&e_{-1,-2}\otimes e^{-\frac{1}{2}%
\sigma }+1\otimes e_{-1,-2}-h_{0,1}\otimes e_{1,1}e^{-\sigma }
 \cr
&&-\widetilde{\widetilde{e_{0,1}}}\otimes e_{1,0}e^{-\sigma }+e_{1,2}\otimes
\widetilde{\widetilde{e_{0,-1}}}e^{-\frac{1}{2}\sigma }\, ,
  \cr
\Delta _{EJ}\left( \widetilde{\widetilde{e_{-1,-1}}}\right)  &=&\widetilde{%
\widetilde{e_{-1,-1}}}\otimes e^{-\frac{1}{2}\sigma }+1\otimes \widetilde{%
\widetilde{e_{-1,-1}}}+e_{1,2}\otimes h_{0,1}e^{-\frac{1}{2}\sigma }
\cr
&&-e_{1,3}\otimes \left( \widetilde{\widetilde{e_{0,-1}}}\right) e^{-\frac{1%
}{2}\sigma }+\frac{4}{3}\left( \widetilde{\widetilde{e_{0,1}}}\right)
\otimes e_{1,1}e^{-\sigma }\, ,
  \cr
\Delta _{EJ}\left( \widetilde{\widetilde{e_{-1,0}}}\right)  &=&\widetilde{%
\widetilde{e_{-1,0}}}\otimes e^{-\frac{1}{2}\sigma }+1\otimes \widetilde{%
\widetilde{e_{-1,0}}}-e_{1,3}\otimes 3h_{0,1}e^{-\frac{1}{2}\sigma }
\cr
&&+e_{1,2}\otimes \widetilde{\widetilde{e_{0,1}}}e^{-\frac{1}{2}\sigma }\, ,
\cr
\Delta _{EJ}\left( \widetilde{\widetilde{e_{-2,-3}}}\right)  &=&\widetilde{%
\widetilde{e_{-2,-3}}}\otimes e^{-\sigma }+1\otimes \widetilde{\widetilde{%
e_{-2,-3}}}
\cr
&&-\widetilde{\widetilde{e_{-1,0}}}\otimes e_{1,0}e^{-\frac{3}{2}\sigma
}+e_{1,3}\otimes \widetilde{\widetilde{e_{-1,-3}}}e^{-\frac{1}{2}\sigma }
\cr
&&-\widetilde{\widetilde{e_{-1,-1}}}\otimes e_{1,1}e^{-\frac{3}{2}\sigma }-%
\frac{2}{3}\widetilde{\widetilde{e_{0,1}}}\otimes \left( e_{1,1}\right)
^{2}e^{-2\sigma }
 \cr
&&+e_{1,2}\otimes e_{-1,-2}e^{-\frac{1}{2}\sigma }+\frac{1}{2}\left(
e_{1,2}\right) ^{2}\otimes \widetilde{\widetilde{e_{0,-1}}}e^{-\sigma }\, .
\cr &&
\end{eqnarray}

These coproducts correspond to the adjoint action of the algebra
${G}_{EJ}^{c\#}$ on the 8-dimensional space $G_{2}\setminus
\mathfrak{g}_{EJ}^{c}$. One can notice two subrepresentations on the
subspaces spanned by $\left\{ \widetilde{\widetilde{e_{0,\pm 1}}},h\right\} $
and $\left\{ \widetilde{\widetilde{e_{-1,-k}}},\widetilde{\widetilde{%
e_{-2,-3}}}\right\} $. On the subspace with the generators $\left\{
\widetilde{\widetilde{e_{0,\pm 1}}},h\right\} $ we have the trivial
factor-representation. This property
  means that on the plane orthogonal to the
initial root we have the $sl(2)$ subalgebra with primitive generators $%
\left\{ \widetilde{\widetilde{e_{0,\pm 1}}},h\right\} $. Such an
effect was first described in \cite{KLO}.

In the deformation $U\left(
g_{2}\right) \longrightarrow U_{EJ}\left( g_{2}\right) $ the costructure (%
\ref{co-car-e1}-\ref{co-car-e2}) on the carrier subalgebra  is
  the extended
Jordanian twist with two extension factors
 (as can be seen for example in $U_{EJ}\left(
sl(4)\right) $). The specific properties of $g_{2}$ become important  in the
negative sector (relations (\ref{new-copr-2})) where the peculiarities of
the root system induce additional terms. For example, the coproducts $\Delta
_{EJ}\left( \widetilde{\widetilde{e_{-1,-1}}}\right) $ and $\Delta
_{EJ}\left( \widetilde{\widetilde{e_{-1,-2}}}\right) $ contain the last two
terms
 depending on the  generators $%
\widetilde{\widetilde{e_{0,1}}}$ and $\widetilde{\widetilde{e_{0,-1}}}$.

Comparing the generators $\widetilde{\widetilde{e}}_{k,l}$
  in (\ref{new-coord-3})%
 with $\widetilde{e}_{k,l}$ (see (4.9) and (4.15)) one can see that most of
 them  have different expressions in terms of the initial $\mathfrak{g}$-basis.
The reason is that the twisting (\ref{full-ext-tw}) not only deforms the group $%
G_{E_{+}J}^{\#}$ %\rightarrow G_{EJ}^{\#}$
but also changes its realization in
terms of the initial $g_{2}$ generators.

\subsection{The Full Chain of Twists for $G_{2}$: Adding the Second Jordanian
 Twist}%%%%%%%%%%%czy male gggggggggggggggggggggggg

The existence of the subalgebra $U\left( sl(2)\right) \subset
U_{EJ}\left( g_{2}\right) $ with primitive generators $\left\{ \widetilde{%
\widetilde{e_{0,\pm 1}}},h\right\} $ shows that the Hopf algebra $%
U_{EJ}\left( g_{2}\right) $ can be additionally deformed
 by the second Jordanian twist. In other words the twist
equations  (\ref{mb3.1}-\ref{mb3.2})
 with $\Delta^{(0)}$ replaced by  $\Delta _{EJ}$ have the
   solution ${{{\cal F}}}_{J^{\prime
}}=e^{h\otimes \omega }$,
 where $\omega$ is given by (\ref{ws3.17}), i.e. one can perform
  the transformation $%
{{{\cal F}}}_{J^{\prime }}:U_{EJ}\left( g_{2}\right) \longrightarrow U_{J^{\prime
}EJ}\left( g_{2}\right) :=U_{C}\left( g_{2}\right) $.
The same result can be achieved  by the chain of twists (\ref{ll2.1}):%
\begin{equation}
{{{\cal F}}}_{C}={{{\cal F}}}_{J^{\prime }EJ}=e^{h\otimes \omega }e^{-e_{1,2}\otimes
e_{1,1}e^{-\frac{1}{2}\sigma }}e^{e_{1,3}\otimes e_{1,0}e^{-\frac{1}{2}%
\sigma }}e^{H\otimes \sigma }  \label{full-chain}
\end{equation}%
applied to the initial $U\left( g_{2}\right) $.

The carrier subalgebra $\mathfrak{g}%
_{C}^{c}$ is the Borel subalgebra ${%\widehat
{\mathfrak{b}}}^{+}\left( g_{2}\right) $. Applying the
twist (\ref{full-chain})
 to $U(g_{2})$ we
obtain the deformed costructure $\Delta _{J^{\prime }EJ}:=\Delta _{C}$
corresponding to the maximal carrier subalgebra in $g_{2}$. In the Hopf
algebra $U_{C}\left( g_{2}\right) $ we have two $\sigma $ -like generators,%
\begin{eqnarray}
\Delta _{C}\left( \sigma \right)  &=&\sigma \otimes 1+1\otimes \sigma , \cr
\Delta _{C}\left( \omega \right)  &=&\omega \otimes 1+1\otimes \omega ,
\end{eqnarray}%
and the ordinary form of the coproduct for the Cartan generator of
${\mathfrak{b}} sl(2)$
twisted by ${{{\cal F}}}_{J^{\prime }}=e^{h\otimes \omega }$:
\[
\Delta _{C}\left( h\right) =h\otimes e^{-\omega }+1\otimes h.
\]

In the coproducts for the elements corresponding to the positive long
sequence $\alpha_1 +k\alpha_2 $ ($k=0,1,2,3$) one can trace the adjoint action of the algebra $%
\mathfrak{g}_{C}^{\#}$:

\begin{eqnarray}
\Delta _{C}\left( e_{1,3}\right)  &=&e_{1,3}\otimes e^{-\frac{1}{2}\sigma +%
\frac{3}{2}\omega }+1\otimes e_{1,3}\, ,
 \cr
\Delta _{C}\left( e_{1,2}\right)  &=&e_{1,2}\otimes e^{-\frac{1}{2}\sigma +%
\frac{1}{2}\omega }+1\otimes e_{1,2}-h\otimes e_{1,3}e^{-\omega }\, ,
 \cr
\Delta _{C}\left( e_{1,1}\right)  &=&e_{1,1}\otimes e^{\frac{1}{2}\sigma -%
\frac{1}{2}\omega }+e^{\sigma }\otimes e_{1,1}
\cr
&&-he^{\sigma }\otimes \left( e_{1,2}e^{-\omega }+\frac{1}{2}%
e_{1,3}e^{-2\omega }\right)
 \cr
&&+\frac{1}{2}h^{2}e^{\sigma }\otimes e_{1,3}e^{-2\omega }\, , \\
\Delta _{C}\left( e_{1,0}\right)  &=&e_{1,0}\otimes e^{\frac{1}{2}\sigma %
\frac{3}{2}\omega }+e^{\sigma }\otimes e_{1,0}
\cr
&&-he^{\sigma }\otimes \left( e_{1,1}e^{-\omega }+\frac{1}{12}%
e_{1,3}e^{-3\omega }+\frac{1}{2}e_{1,2}e^{-2\omega }\right)
\cr
&&+\frac{1}{2}h^{2}e^{\sigma }\otimes \left( e_{1,2}e^{-2\omega
}+e_{1,3}e^{-3\omega }\right)
 \cr
&&-\frac{1}{6}h^{3}e^{\sigma }\otimes e_{1,3}e^{-3\omega }\, .
\end{eqnarray}%
The terms corresponding to the powers of $\mathrm{ad}\left( X_{h}\right) $
of the short root operator dual to $h$ (such as $-\frac{1}{6}h^{3}e^{\sigma
}\otimes e_{1,3}e^{-3\omega }$ in the last row) are accompanied by the
additional summands that will
 disappear when we  pass (via the second classical limit)
  to the group costructure for $G_{C}^{\#}$. This is
especially evident when $\Delta _{C}\left( H\right) $ is considered:%
\begin{eqnarray}
\label{lutop5.14}
\Delta _{C}\left( H\right)  &=&e^{\mathrm{ad}h\otimes \omega }\left(
\begin{array}{c}
H\otimes e^{-\sigma }+1\otimes H \cr
-e_{1,3}\otimes e_{1,0}e^{-\frac{3}{2}\sigma }+e_{1,2}\otimes e_{1,1}e^{-%
\frac{3}{2}\sigma }%
\end{array}%
\right)  \cr
&=&H\otimes e^{-\sigma }+1\otimes H
 \cr
&&
+h\otimes \left( -\frac{1}{2}\left( e_{1,2}\right) ^{2}e^{-\omega }-\frac{1%
}{2}e_{1,2}e_{1,3}e^{-2\omega }-\frac{1}{3}\left( e_{1,3}\right)
^{2}e^{-3\omega }\right)
 \cr
&&+\frac{1}{2}h^{2}\otimes \left( e_{1,3}e_{1,2}e^{-2\omega }+\left(
e_{1,3}\right) ^{2}e^{-3\omega }\right)
\cr
&&+\frac{1}{6}h^{3}\otimes \left( -\left( e_{1,3}\right) ^{2}e^{-3\omega
}\right) -e_{1,3}\otimes e_{1,0}e^{-\frac{3}{2}\left( \sigma -\omega \right)
}
\cr
&&-he_{1,3}\otimes \left( -e_{1,1}e^{-\frac{5}{2}\sigma +\frac{3}{2}\omega }-%
\frac{1}{12}e_{1,3}e^{-\frac{9}{2}\sigma +\frac{3}{2}\omega }-\frac{1}{2}%
e_{1,2}e^{-\frac{7}{2}\sigma +\frac{3}{2}\omega }\right)
\cr
&&-\frac{1}{2}h^{2}e_{1,3}\otimes \left( e_{1,2}e^{-\frac{7}{2}\sigma +\frac{%
3}{2}\omega }+e_{1,3}e^{-\frac{9}{2}\sigma +\frac{3}{2}\omega }\right)
\cr
&&+ \, \frac{1}{6}h^{3}e_{1,3}\otimes e_{1,3}e^{-\frac{9}{2}+ \sigma \frac{%
3}{2}\omega }
\cr
&&+e_{1,2}\otimes e_{1,1}e^{-\frac{3}{2}\sigma +\frac{1}{2}\omega
}-he_{1,2}\otimes e_{1,2}e^{-\frac{3}{2}\sigma -\frac{1}{2}\omega }
\cr
&&-\frac{1}{2}he_{1,2}\otimes e_{1,3}e^{-\frac{3}{2}\sigma -\frac{3}{2}%
\omega }+\frac{1}{2}h^{2}e_{1,2}\otimes e_{1,3}e^{-\frac{3}{2}\sigma -\frac{3%
}{2}\omega }\, .
\nonumber \\
\end{eqnarray}

One can show that in the second classical limit the
 large number  of terms in (\ref{lutop5.14}) will disappear.

Let us turn now to the determination of $\mathfrak{g}_{C}^{\#}$- basis
in $U_{C}(g_2)$.

In the negative root sector ${\mathfrak{n}}_{-}$ the element
 $\widetilde{\widetilde{e_{0,-1}}}$ must be evidently changed:
\begin{equation}
\widehat{e_{0,-1}}=\widetilde{\widetilde{e_{0,-1}}}-\frac{1}{3}h^{2}.
\end{equation}
This is due to the fact that on the subspace generated by $\widetilde{%
\widetilde{e_{0,1}}}$, $\widetilde{\widetilde{e_{0,-1}}}$ and $h$ the twist
deformation performed by ${{{\cal F}}}_{J^{\prime }}$ is an ordinary Jordanian
deformation for the algebra $sl(2)$. The coproduct for $\widehat{e_{0,-1}}$
is quasiprimitive:
\begin{equation}
\Delta _{C}\left( \widehat{e_{0,-1}}\right) =\widehat{e_{0,-1}}\otimes
e^{-\omega }+1\otimes \widehat{e_{0,-1}}.  \label{xminus}
\end{equation}

The other coproducts for negative sector generators
  are  quite complicated  despite the fact that they
are obtained in the improved (dual group $G_{EJ}^{\#}$) basis $\left\{
\widetilde{\widetilde{x}}\right\}$.  The factor ${{{\cal F}}}_{J^{\prime }}$
 changes  the dual group
\begin{equation}
{{{\cal F}}}_{J^{\prime }}:G_{EJ}^{\#}\longrightarrow G_{C}^{\#}
\end{equation}
and the corresponding Lie algebra generators must be redefined. Using the
technique demonstrated in Section~III.D we get the set of
 new generators:

\bl
\begin{eqnarray}
\widehat{e_{0,1}} &=&\widetilde{\widetilde{e_{0,1}}}=e_{0,1}+\frac{1}{2}%
\left( e_{1,2}\right) ^{2}=e^{\omega }-1;
\\
\widehat{e_{0,-1}} &=&\widetilde{\widetilde{e_{0,-1}}}-\frac{1}{3}%
h^{2}=e_{0,-1}+\frac{2}{3}\left( e_{1,1}\right) ^{2}e^{-\sigma }-\frac{1}{3}%
h^{2};
\\
\widehat{e_{-1,0}} &=&\widetilde{\widetilde{e_{-1,0}}}%
+e_{1,2}+he_{1,3}=e_{-1,0}+\left( h-H\right) e_{1,3}
%\nonumber \\
\cr
&&
+\ \frac{1}{6}\left(
e_{1,2}\right) ^{3}+e_{1,2};
\\
\widehat{e_{-1,-1}} &=&\widetilde{\widetilde{e_{-1,-1}}}-\frac{4}{3}%
e_{1,1}e^{-\sigma +\omega }-\frac{1}{3}he_{1,2}
+\frac{1}{3}h^{2}e_{1,3}
%\notag \\
\cr
&=&e_{-1,-1}+\left( H-\frac{1}{3}h\right) e_{1,2}-\frac{4}{3}%
e_{1,1}e^{-\sigma +\omega }+\frac{1}{3}h^{2}e_{1,3};
\\
\widehat{e_{-1,-2}} &=&\widetilde{\widetilde{e_{-1,-2}}}+e_{1,0}e^{-\sigma
+\omega }-\frac{1}{3}h^{2}e_{1,2}
%\notag \\
\cr
&=&e_{-1,-2}+e_{1,0}e^{-\sigma +\omega }-\frac{1}{3}h^{2}e_{1,2};
\\
\widehat{e_{-1,-3}} &=&\widetilde{\widetilde{e_{-1,-3}}}=e_{-1,-3}+\frac{2}{9%
}\left( e_{1,1}\right) ^{3}e^{-2\sigma };
\\
\widehat{e_{-2,-3}} &=&\widetilde{\widetilde{e_{-2,-3}}}+\frac{2}{3}\left(
e_{1,1}\right) ^{2}e^{-2\sigma +\omega }-\frac{1}{6}h^{2}\left(
e_{1,2}\right) ^{2}-\frac{2}{3}h
% \notag \\
\cr
&=&e_{-2,-3}-H^{2}+\frac{2}{3}\left( e_{1,1}\right) ^{2}e^{-2\sigma +\omega
}-\frac{1}{6}h^{2}\left( e_{1,2}\right) ^{2}-\frac{2}{3}h;
\end{eqnarray}
\el

In the new basis we find
 fewer nonzero costructure constants. One
of the new generators corresponds to the
 boundary of ${\mathfrak{n}}_{-}$ and thus is
quasiprimitive,
\begin{equation}
\Delta _{C}\left( \widehat{e_{-1,0}}\right) =\widehat{e_{-1,0}}\otimes e^{-%
\frac{1}{2}\sigma +\frac{3}{2}\omega }+1\otimes \widehat{e_{-1,0}}
\end{equation}
(notice that the other boundary element
 is $\widehat{e_{0,-1}}$ with the
 coproduct (\ref{xminus})).

The number of independent terms in the coproducts corresponds to the number
of different decompositions of the vector
 $\beta \in \sum _{{\mathfrak{n}}_{-}}$ in terms of the $%
\mathfrak{g}_{C}^{c\#}$ roots. This explains
  why in the sequence $e_{-\alpha_1
-k\alpha_2 }$ ($k=0,1,2,3$) the number of terms rapidly
 increases with $k$:

\bl
\begin{eqnarray}
&&\Delta _{C}\left( \widehat{e_{-1,-1}}\right)
  = \widehat{e_{-1,-1}}\otimes
e^{-\frac{1}{2}\sigma +\frac{1}{2}\omega }+1\otimes \widehat{e_{-1,-1}}
\notag
 \\
   \qquad
&&
\qquad
+h\otimes \left( \widehat{e_{-1,0}}+e_{1,3}\right) e^{-\omega
}-e_{1,3}\otimes \widehat{e_{0,-1}}e^{-\frac{1}{2}\sigma +\frac{3}{2}\omega
};
%%\cr &&
\label{delta-1-1}
\cr &&
\\
\cr
&&
\Delta _{C}\left( \widehat{e_{-1,-2}}\right)  =\widehat{e_{-1,-2}}\otimes
e^{-\frac{1}{2}\sigma -\frac{1}{2}\omega }+1\otimes \widehat{e_{-1,-2}}
%\notag \\
\cr\cr
\qquad
&&
\qquad
+h\otimes \widehat{e_{-1,-1}}e^{-\omega }+\frac{1}{2}h^{2}\otimes \widehat{%
e_{-1,0}}e^{-2\omega }+e_{1,2}\otimes \widehat{e_{0,-1}}e^{-\frac{1}{2}%
\sigma +\frac{1}{2}\omega }
%  \notag  \\
\cr\cr
\qquad
&&
\qquad
-\frac{1}{2}h\otimes \widehat{e_{-1,0}}e^{-2\omega }-\frac{1}{3}h\otimes
e_{1,2}e^{-\omega }-\frac{2}{3}\left( h-h^{2}\right) \otimes
e_{1,3}e^{-2\omega };
%%\cr &&
\label{delta-1-2}
\cr &&
 \\
&&
\Delta _{C}\left( \widehat{e_{-1,-3}}\right)
=\widehat{e_{-1,-3}}\otimes
e^{-\frac{1}{2}\sigma -\frac{3}{2}\omega }+1\otimes \widehat{e_{-1,-3}}
\cr
\cr
\qquad
&&
\qquad
+h\otimes \left(
\begin{array}{c}
\widehat{e_{-1,-2}}e^{-\omega }+\frac{1}{3}\widehat{e_{-1,0}}e^{-3\omega }-%
\widehat{e_{0,-1}}e_{1,2}e^{-\omega }-\frac{4}{3}e_{1,1}e^{-\sigma -\omega }
\\
-\frac{1}{2}\widehat{e_{-1,-1}}e^{-2\omega }-\frac{1}{2}e_{1,2}e^{-2\omega }-%
\frac{1}{2}\widehat{e_{0,-1}}e_{1,3}e^{-2\omega }+
\end{array}
\right)
%%   \notag    \\
\cr\cr
\qquad
&&
\qquad
+h^{2}\otimes \left(
\begin{array}{c}
\frac{1}{2}\widehat{e_{-1,-1}}e^{-2\omega }+\frac{1}{2}\widehat{e_{0,-1}}%
e_{1,3}e^{-2\omega }+ \\
+\frac{1}{6}e_{1,2}e^{-2\omega }-\frac{1}{3}e_{1,3}e^{-3\omega }-\frac{1}{2}%
\widehat{e_{-1,0}}e^{-3\omega }
\end{array}
\right)
   \notag
   \\
\qquad
&&
\qquad
+h^{3}\otimes \left( \frac{1}{6}\widehat{e_{-1,0}}e^{-3\omega }+\frac{1}{3}%
e_{1,3}e^{-3\omega }\right)
\notag
 \\
\qquad
&&
\qquad
+h\widehat{e_{0,-1}}\otimes \left( -e_{1,2}e^{-2\omega }-\frac{1}{2}%
e_{1,3}e^{-3\omega }\right)
   \notag
    \\
\qquad
&&
\qquad
+\widehat{e_{0,-1}}\otimes e_{1,1}e^{-\sigma -\omega }+\frac{1}{2}h^{2}%
\widehat{e_{0,-1}}\otimes e_{1,3}e^{-3\omega };  \label{delta-1-3}
\cr &&
\end{eqnarray}
\el

In the twisted algebra $U_{C}(g_{2})$ the group multiplication of $G_{C}^{\#}
$ is nontrivially deformed.
In particular the remaining coproduct $\Delta _{C}\left(
\widehat{e_{-2,-3}}\right) $
 is the largest one: it  has 23 terms that correspond to the adjoint action
of the dual carrier group $G_{C}^{c\#}$ on the space
 of its 6-dimensional subrepresentation, and the other
 $28$ terms do   appear only due to the noncommutativity of the
coordinates.

%%%%dalszy ciag 29 29 29
\begin{eqnarray}
%\begin{array}{lll}
&&
\Delta _{C}\left( \widehat{e_{-2,-3}}\right)
 = \widehat{e_{-2,-3}}\otimes
e^{-\sigma }+1\otimes \widehat{e_{-2,-3}}
%\nonumber \\[12pt]
\cr\cr
&&
\qquad
-\widehat{e_{-1,0}}\otimes e_{1,0}e^{-\frac{3}{2}\sigma +\frac{3}{2}\omega
}-\widehat{e_{-1,-1}}\otimes e_{1,1}e^{-\frac{3}{2}\sigma +\frac{1}{2}\omega
}
%\nonumber \\[12pt]
\cr\cr
&&
\qquad
+e_{1,2}\otimes \widehat{e_{-1,-2}}e^{-\frac{1}{2}\sigma +\frac{1}{2}%
\omega }+e_{1,3}\otimes \widetilde{\widetilde{e_{-1,-3}}}e^{-\frac{1}{2}%
\sigma +\frac{3}{2}\omega }
%\nonumber \\[12pt]
%\cr \cr
\nonumber \\
\cr
&&
\qquad
+\frac{1}{2}\left( e_{1,2}\right) ^{2}\otimes \widehat{e_{0,-1}}e^{-\sigma
+\omega }
%\nonumber \\[12pt]
%\cr\cr
\nonumber \\
\cr
&&
\qquad
+h\otimes \widehat{e_{-1,-1}}e_{1,2}e^{-\omega }-\frac{1}{2}h^{2}\otimes
\widehat{e_{-1,-1}}e_{1,3}e^{-2\omega }
%\nonumber \\[12pt]
\cr\cr
&&
\qquad
+he_{1,2}\otimes \widehat{e_{-1,-1}}e^{-\frac{1}{2}\sigma -\frac{1}{2}%
\omega }+h\widehat{e_{-1,-1}}\otimes e_{1,2}e^{-\frac{1}{2}\sigma -
\frac{1}{2%
}\omega }
%\nonumber \\[12pt]
\cr\cr
&&
\qquad
+\frac{1}{2}h^{2}e_{1,3}\otimes \widehat{e_{-1,-1}}e^{-\frac{1}{2}\sigma -%
\frac{1}{2}\omega }-\frac{1}{2}h^{2}\widehat{e_{-1,-1}}\otimes e_{1,3}e^{-%
\frac{1}{2}\sigma -\frac{3}{2}\omega }
%\nonumber \\[12pt]
\cr
\cr
&&
\qquad
+\frac{1}{2}h^{2}\otimes \widehat{e_{-1,0}}e_{1,2}e^{-2\omega
}+\frac{1}{2}%
h^{2}e_{1,3}\otimes \widehat{e_{0,-1}}e_{1,3}e^{-\frac{1}{2}\sigma -
\frac{1}{%
2}\omega }
%\nonumber \\[12pt]
\cr
\cr
&&
\qquad
-\frac{1}{3}h^{3}\otimes \widehat{e_{-1,0}}e_{1,3}e^{-3\omega
}+h\widehat{%
e_{-1,0}}\otimes e_{1,1}e^{-\frac{3}{2}\sigma +\frac{1}{2}\omega }
%\nonumber \\[12pt]
\cr
\cr
&&
\qquad
-\frac{1}{2}h^{2}\widehat{e_{-1,0}}\otimes e_{1,2}e^{-\frac{1}{2}\sigma +%
\frac{1}{2}\omega }+\frac{1}{2}h^{2}e_{1,2}\otimes \widehat{e_{-1,0}}e^{-%
\frac{1}{2}\sigma -\frac{3}{2}\omega }
%\nonumber \\[12pt]
\cr
\cr
&&
\qquad
+\frac{1}{6}h^{3}e_{1,3}\otimes \widehat{e_{-1,0}}e^{-\frac{1}{2}\sigma -%
\frac{3}{2}\omega }+\frac{1}{6}h^{3}\widehat{e_{-1,0}}\otimes e_{1,3}e^{-%
\frac{1}{2}\sigma -\frac{3}{2}\omega }
%\nonumber \\[12pt]
\cr
\cr
&&
\qquad
+ he_{1,3}\otimes \widehat{e_{-1,-2}}e^{-\frac{1}{2}\sigma +\frac{1}{2}%
\omega }-he_{1,3}\otimes \widehat{e_{0,-1}}e_{1,2}e^{-\frac{1}{2}\sigma +%
\frac{1}{2}\omega }
%%%%%%%%%%%%%%%%%%%%%%%%%%%%%%%%%%%%%%
%%%%%%%%%%%%%%%%%%%%\end{eqnarray*}
 % \end{array}
\nonumber \\
%\end{eqnarray}
%
%
%\begin{eqnarray}
\cr
%%%%%%%%%%%%%%%%%%%\begin{array}{rll}
%%%%%%%%%%%%%%%%%%%%%%%%%%%%%%%%%%%%%%
%%%%%%%%%%%%%%\begin{eqnarray}
&&
\qquad
-h\otimes \left(
\begin{array}{c}
\frac{4}{3}e_{1,1}e_{1,3}e^{-\sigma -\omega }+\frac{3}{2}e_{1,2}e_{1,3}e^{-
2%
\omega }+\frac{1}{3}\left( e_{1,2}\right) ^{2}e^{-\omega }
\\[6pt]
+\frac{4}{3}\left( e_{1,3}\right) ^{2}e^{-3\omega }+\frac{1}{2}\widehat{%
e_{-1,0}}e_{1,2}e^{-2\omega }
 \\[6pt]
+\frac{2}{3}\widehat{e_{-1,0}}e_{1,3}e^{-3\omega }-\frac{1}{2}\widehat{%
e_{-1,-1}}e_{1,3}e^{-2\omega }
\end{array}
\right)
  \notag \\
  \cr
&&
\qquad
-he_{1,2}\otimes \left( \frac{2}{3}e_{1,2}e^{-\frac{1}{2}\sigma -\frac{1}{2%
}\omega }+\frac{5}{12}e_{1,3}e^{-\frac{1}{2}\sigma -\frac{3}{2}\omega
}\right)
  \notag \\
  \cr
&&
\qquad
-he_{1,3}\otimes \left(
\begin{array}{c}
\frac{4}{3}e_{1,1}e^{-\frac{3}{2}\sigma +\frac{1}{2}\omega }+\frac{1}{2}%
e_{1,2}e^{-\frac{1}{2}\sigma -\frac{1}{2}\omega }
 \\[6pt]
-\frac{1}{3}\widehat{e_{-1,0}}e^{-\frac{1}{2}\sigma -\frac{3}{2}\omega }+%
\frac{1}{2}\widehat{e_{0,-1}}e_{1,3}e^{-\frac{1}{2}\sigma -\frac{1}{2}\omega
}
 \\[6pt]
+\frac{1}{2}\widehat{e_{-1,-1}}e^{-\frac{1}{2}\sigma -\frac{1}{2}\omega }
\end{array}
\right)
 \notag
 \\
 \cr
&&
\qquad
+h^{2}\otimes \left( \frac{1}{2}e_{1,2}e_{1,3}e^{-2\omega }+2\left(
e_{1,3}\right) ^{2}e^{-3\omega }+\widehat{e_{-1,0}}e_{1,3}e^{-3\omega
}\right)
   \notag \\
   \cr
&&
\qquad
+h^{2}e_{1,2}\otimes \left( \frac{1}{6}e_{1,3}e^{-\frac{1}{2}\sigma -\frac{%
3}{2}\omega }+\frac{1}{2}\widehat{e_{-1,0}}e^{-\frac{1}{2}\sigma -
\frac{3}{2}%
\omega }\right)
 \notag \\
 \cr
&&
\qquad
+h^{2}e_{1,3}\otimes \left( \frac{1}{6}e_{1,2}e^{-\frac{1}{2}\sigma -\frac{%
1}{2}\omega }-\frac{1}{3}e_{1,3}e^{-\frac{1}{2}\sigma -\frac{3}{2}\omega }-%
\frac{1}{2}\widehat{e_{-1,0}}e^{-\frac{1}{2}\sigma -\frac{3}{2}\omega
}\right)
  \notag \\
  \cr
&&
\qquad
-\frac{1}{2}h^{3}\otimes \left( e_{1,3}\right) ^{2}e^{-3\omega }+\frac{1}{3%
}h^{3}e_{1,3}\otimes e_{1,3}e^{-\frac{1}{2}\sigma -\frac{3}{2}\omega }
\notag \\
\cr
&&
\qquad
+h\widehat{e_{-1,0}}\otimes \left( \frac{1}{3}e_{1,3}e^{-\frac{1}{2}\sigma
-\frac{3}{2}\omega }+\frac{1}{2}e_{1,2}e^{-\frac{1}{2}\sigma -\frac{1}{2}%
\omega }\right)
\notag
 \\
&&
\qquad
+\frac{1}{2}h\widehat{e_{-1,-1}}\otimes e_{1,3}e^{-\frac{1}{2}\sigma -%
\frac{3}{2}\omega }-\frac{1}{2}h^{2}\widehat{e_{-1,0}}\otimes e_{1,3}e^{-%
\frac{1}{2}\sigma -\frac{3}{2}\omega }\, .
\label{delta-2-3}
%%%%%%%%%%%%%%%%%% \end{array}
\end{eqnarray}

%%%%!!!!!!!!!!!!!!!!!!!!!!!!!!!!!!!!!!!!!!!!

Comparing these results with those for
   other simple Lie algebras (deformed
by the full chains of twists) we see that contrary
 e.g.  to the $sl(n)$ case
in the twisted algebra $U_{C}(G_{2})$ the dual
  generator $\widehat{e_{-2,-3}%
}$ differs considerably from the ''Jordanian'' $\widetilde{\widetilde{%
e_{-2,-3}}}=e_{-2,-3}-H^{2}$ and
 as well  most of the coproducts in the negative
sector are strongly deformed. In this situation the appropriate basis for
the presentation of the costructure plays a very essential  role.
 In the initial $%
\mathfrak{g}$-basis the decomposition of the coproduct $\Delta _{C}\left(
e_{-2,-3}\right) $ contains more than four hundred terms. Using the $%
\mathfrak{g}_{EJ}^{\#}$ -basis $\left\{ \widetilde{\widetilde{x}}\right\} $
we can reduce this number to 109 and finally
 in $\mathfrak{g}_{C}^{\#}$-basis
we get the expression (\ref{delta-2-3}) with 51 terms. The effective
technique is needed to perform the corresponding calculations and this is
where the modified BCH-formulas (\ref{a3.30}--\ref{a3.31}) presented in
Subsection~III.D
 are very useful. The expressions (\ref{delta-1-1}--\ref{delta-1-3})
  and (\ref{delta-2-3}) were obtained with their help.
%%%

The full chain of twists (\ref{full-chain}) can be parametrized as
 follows:
\begin{equation}
\label{blm5.22}
{{{\cal F}}}_{C}={{{\cal F}}}_{J^{\prime }}{{{\cal F}}}_{E}{{{\cal F}}}_{J}=e^{h\otimes \omega \left( \xi ,\psi \right)
}e^{-\xi e_{1,2}\otimes e_{1,1}e^{-\frac{1}{2}\sigma \left( \xi \right)
}}e^{\xi e_{1,3}\otimes e_{1,0}e^{-\frac{1}{2}\sigma \left( \xi \right)
}}e^{H\otimes \sigma \left( \xi \right) }
\end{equation}
with
\begin{eqnarray}
\sigma \left( \xi \right)  &=&\ln \left( 1+\xi e_{2,3}\right) ,
\cr
\omega \left( \xi ,\psi \right)  &=&\ln \left( 1+\psi e_{0,1}+\frac{1}{2}\xi
\psi \left( e_{1,2}\right) ^{2}\right) \, .
\end{eqnarray}
 Thus the full chain leads to the 2-parameter set of Hopf algebras $%
U_{C}(G_{2};\xi , \psi )$. In particular this parametrization provides
the possibility  to study the second classical limit for $U_{C}(G_{2})$.
    The latter is obtained
by scaling the generators $x\longrightarrow \frac{1}{\varepsilon }x$ and by
going to the limit $\varepsilon ,\xi ,\psi \longrightarrow 0,$ with
 finite values of
  $\frac{\xi }{\varepsilon }=\zeta ,\frac{\psi }{\varepsilon }=\eta$.
   In such a
   limit we get the composition law of the dual group $G_{C}^{\#}$:

\bl
\begin{eqnarray}
%\begin{array}{rll}
\label{lutop55a}
&& \Delta^{\#}_C \left( H\right) =
 H\otimes e^{-\sigma \left( \zeta \right)
}+1\otimes H
\nonumber \\
&&
\qquad
-\zeta e_{1,3}
 \otimes e_{1,0}e^{-\frac{3}{2}
 \left( \sigma \left( \zeta
\right)
-\omega \left( \zeta , \eta \right) \right) }+\zeta e_{1,2}\otimes
e_{1,1}e^{-\frac{3}{2}\sigma \left( \zeta \right) +\frac{1}{2}\omega \left( \zeta , \eta \right) }
\nonumber \\
&&
\qquad
-\frac{1}{2}\zeta \eta h\otimes \left( e_{1,2}\right) ^{2}e^{-\omega \left( \zeta , \eta \right) }-\zeta \eta he_{1,2}\otimes e_{1,2}e^{-\frac{3}{2}\sigma
\left( \zeta \right) -\frac{1}{2}\omega \left( \zeta , \eta \right) }
\nonumber \\
&&
\qquad
+\zeta \eta he_{1,3}\otimes e_{1,1}e^{-\frac{5}{2}\sigma \left( \zeta
\right) +\frac{3}{2}\omega \left( \zeta , \eta \right) }+\frac{1}{2}\zeta \eta
^{2}h^{2}e_{1,2}\otimes e_{1,3}e^{-\frac{3}{2}\sigma \left( \zeta \right) -%
\frac{3}{2}\omega \left( \zeta , \eta \right) }
\nonumber \\
&&
\qquad
+\frac{1}{2}\zeta \eta ^{2}h^{2}\otimes e_{1,3}e_{1,2}e^{-2\omega \left( \zeta , \eta \right) }-\frac{1}{2}\zeta \eta ^{2}h^{2}e_{1,3}\otimes
e_{1,2}e^{-\frac{7}{2}\sigma \left( \zeta \right) +\frac{3}{2}\omega \left( \zeta , \eta \right) }
\nonumber \\
&&
\qquad
-\frac{1}{6}\zeta \eta ^{3}h^{3}\otimes \left( e_{1,3}\right)
^{2}e^{-3\omega \left( \zeta , \eta \right) }+\frac{1}{6}\zeta \eta
^{3}h^{3}e_{1,3}\otimes e_{1,3}e^{-\frac{9}{2}\sigma \left( \zeta \right)
\frac{3}{2}\omega \left( \zeta , \eta \right) }\, ,
%\end{array}
%\nonumber
\cr &&
\end{eqnarray}
\begin{eqnarray}
 \Delta^{\#}_C \left( h\right)
 & = &
 h\otimes e^{-\omega \left( \zeta , \eta \right) }+1\otimes h \, ,
\\
 \Delta^{\#}_C \left( e_{1,0}\right)
 & = &
 e_{1,0}\otimes e^{\frac{1}{2}\sigma
\left( \zeta \right) -\frac{3}{2}\omega \left( \zeta , \eta \right)
}+e^{\sigma \left( \zeta \right) }\otimes e_{1,0}
\nonumber \\
&&
-\eta he^{\sigma \left( \zeta \right) }\otimes e_{1,1}e^{-\omega \left( \zeta , \eta \right) }+\frac{1}{2}\eta ^{2}h^{2}e^{\sigma \left( \zeta \right)
}\otimes e_{1,2}e^{\sigma \left( \zeta \right) -2\omega \left( \zeta , \eta \right) }
\nonumber \\
&&
-\frac{1}{6}\eta ^{3}h^{3}e^{\sigma \left( \zeta \right) }\otimes
e_{1,3}e^{\sigma \left( \zeta \right)
 -3\omega \left( \zeta , \eta \right) }\, ,
\\
 \Delta^{\#}_C \left( e_{1,1}\right)
 & = &
  e_{1,1}\otimes e^{\frac{1}{2}\sigma
\left( \zeta \right) -\frac{1}{2}\omega \left( \zeta , \eta \right)
}+e^{\sigma \left( \zeta \right) }\otimes e_{1,1}
\cr
&&
-\eta he^{\sigma \left( \zeta \right) }\otimes e_{1,2}e^{\sigma \left( \zeta
\right) -\omega \left( \zeta , \eta \right) }
\cr
&&
+\frac{1}{2}\eta
^{2}h^{2}e^{\sigma \left( \zeta \right) }\otimes e_{1,3}e^{\sigma \left(
\zeta \right) -2\omega \left( \zeta , \eta \right) }
\cr && \\
 \Delta^{\#}_C \left( e_{1,2}\right)
  & = &
   e_{1,2}\otimes e^{-\frac{1}{2}\sigma
\left( \zeta \right) +\frac{1}{2}\omega \left( \zeta , \eta \right) }+1\otimes
e_{1,2}-\eta h\otimes e_{1,3}e^{-\omega \left( \zeta , \eta \right) }\, ,
\cr && \\
 \Delta^{\#}_C \left( e_{1,3}\right)
 & = &
 e_{1,3}\otimes e^{-\frac{1}{2}\sigma
\left( \zeta \right) +\frac{3}{2}\omega \left( \zeta , \eta \right) }+1\otimes
e_{1,3} \, ,
\cr && \\
 \Delta^{\#}_C \left( e_{2,3}\right)
 &= &
  e_{2,3}\otimes e^{\sigma \left( \zeta
\right) }+1\otimes e_{2,3} \, ,
\cr && \\
 \Delta^{\#}_C \left( \omega \left( \zeta , \eta \right) \right)
 & = &
 \omega \left( \zeta , \eta \right) \otimes 1+1\otimes \omega \left( \zeta , \eta \right) \, ,
\cr && \\
 \Delta^{\#}_C \left( \widehat{e_{0,-1}}\right)
 & = &
 \widehat{e_{0,-1}}\otimes
e^{-\omega \left( \zeta , \eta \right) }+1\otimes \widehat{e_{0,-1}}\, ,
\cr && \\
 \Delta^{\#}_C \left( \widehat{e_{-1,0}}\right)
 &= &
 \widehat{e_{-1,0}}\otimes
e^{-\frac{1}{2}\sigma \left( \zeta \right) +\frac{3}{2}\omega \left( \zeta , \eta \right) }+1\otimes \widehat{e_{-1,0}}\, ,
\cr && \\
 \Delta^{\#}_C \left( \widehat{e_{-1,-1}}\right)
 & = &
 \widehat{e_{-1,-1}}%
\otimes e^{-\frac{1}{2}\sigma \left( \zeta \right) +\frac{1}{2}\omega \left( \zeta , \eta \right) }+1\otimes \widehat{e_{-1,-1}}
\cr
&&
+\eta h\otimes \widehat{e_{-1,0}}e^{-\omega \left( \zeta , \eta \right)
}-\zeta e_{1,3}\otimes \widehat{e_{0,-1}}e^{-\frac{1}{2}\sigma \left( \zeta
\right) +\frac{3}{2}\omega \left( \zeta , \eta \right) }\, ,
\cr && \\
 \Delta^{\#}_C \left( \widehat{e_{-1,-2}}\right)
& =&
  \widehat{e_{-1,-2}}%
\otimes e^{-\frac{1}{2}\sigma \left( \zeta \right) -\frac{1}{2}\omega \left( \zeta , \eta \right) }+1\otimes \widehat{e_{-1,-2}}
\cr
&&
+\eta h\otimes \widehat{e_{-1,-1}}e^{-\omega \left( \zeta , \eta \right) }+%
\frac{1}{2}\eta ^{2}h^{2}\otimes \widehat{e_{-1,0}}e^{-2\omega \left( \zeta , \eta \right) }
\cr
&&
+\zeta e_{1,2}\otimes \widehat{e_{0,-1}}e^{-\frac{1}{2}\sigma \left( \zeta
\right) +\frac{1}{2}\omega \left( \zeta , \eta \right) }\, ,
\cr && %%&&  \\
\end{eqnarray}
\begin{eqnarray}
&& \Delta^{\#}_C \left( \widehat{e_{-1,-3}}\right)
 =
 \widehat{e_{-1,-3}}%
\otimes e^{-\frac{1}{2}\sigma \left( \zeta \right) -\frac{3}{2}\omega \left( \zeta , \eta \right) }+1\otimes \widehat{e_{-1,-3}}
\nonumber \\
&&
+h\otimes \left( \eta \widehat{e_{-1,-2}}e^{-\omega \left( \zeta , \eta \right) }-\zeta \eta \widehat{e_{0,-1}}e_{1,2}e^{-\omega \left( \zeta , \eta \right) }\right)
\nonumber \\
&&
+h^{2}\otimes \left( \frac{1}{2}\eta ^{2}\widehat{e_{-1,-1}}e^{-2\omega \left( \zeta , \eta \right) }+\frac{1}{2}\zeta \eta ^{2}\widehat{e_{0,-1}}%
e_{1,3}e^{-2\omega \left( \zeta , \eta \right) }\right)
\nonumber \\
&&
+\frac{1}{6}\eta ^{3}h^{3}\otimes \widehat{e_{-1,0}}e^{-3\omega \left( \zeta , \eta \right) }-\zeta \eta h\widehat{e_{0,-1}}\otimes e_{1,2}e^{-2\omega \left( \zeta , \eta \right) }
\nonumber \\
&&
+\zeta \widehat{e_{0,-1}}\otimes e_{1,1}e^{-\sigma \left( \zeta \right)
-\omega \left( \zeta , \eta \right) }+\frac{1}{2}\zeta \eta ^{2}h^{2}\widehat{%
e_{0,-1}}\otimes e_{1,3}e^{-3\omega \left( \zeta , \eta \right) }\, ,
\cr && \\
&&
 \Delta^{\#}_C \left( \widehat{e_{-2,-3}}\right)
  =  \widehat{e_{-2,-3}}%
\otimes e^{-\sigma \left( \zeta \right) }+1\otimes \widehat{e_{-2,-3}}
\nonumber \\
&&
\
-\zeta \widehat{e_{-1,0}}\otimes e_{1,0}e^{-\frac{3}{2}\sigma \left( \zeta
\right) +\frac{3}{2}\omega \left( \zeta ,\eta \right) }+\zeta e_{1,3}\otimes
\widehat{e_{-1,-3}}e^{-\frac{1}{2}\sigma \left( \zeta \right) +\frac{3}{2}%
\omega \left( \zeta ,\eta \right) }
\nonumber \\
&&
\
-\zeta \widehat{e_{-1,-1}}\otimes e_{1,1}e^{-\frac{3}{2}\sigma \left( \zeta
\right) +\frac{1}{2}\omega \left( \zeta ,\eta \right) }+\zeta e_{1,2}\otimes
\widehat{e_{-1,-2}}e^{-\frac{1}{2}\sigma \left( \zeta \right) +\frac{1}{2}%
\omega \left( \zeta ,\eta \right) }
\nonumber \\
&&
\
+h\otimes \zeta \eta \widehat{e_{-1,-1}}e_{1,2}e^{-\omega \left( \zeta ,\eta
\right) }-\frac{1}{2}h^{2}\otimes \zeta \eta ^{2}\widehat{e_{-1,-1}}%
e_{1,3}e^{-2\omega \left( \zeta ,\eta \right) }
\nonumber \\
&&
\
+\zeta \eta he_{1,2}\otimes \widehat{e_{-1,-1}}e^{-\frac{1}{2}\sigma \left(
\zeta \right) -\frac{1}{2}\omega \left( \zeta ,\eta \right) }+\zeta \eta h%
\widehat{e_{-1,-1}}\otimes e_{1,2}e^{-\frac{1}{2}\sigma \left( \zeta \right)
-\frac{1}{2}\omega \left( \zeta ,\eta \right) }
\nonumber \\
&&
\
+\frac{1}{2}\zeta \eta ^{2}h^{2}e_{1,3}\otimes \widehat{e_{-1,-1}}e^{-\frac{1%
}{2}\sigma \left( \zeta \right) -\frac{1}{2}\omega \left( \zeta ,\eta
\right) }
\cr
&&
-\frac{1}{2}\zeta \eta ^{2}h^{2}\widehat{e_{-1,-1}}\otimes
e_{1,3}e^{-\frac{1}{2}\sigma \left( \zeta \right) -\frac{3}{2}\omega \left(
\zeta ,\eta \right) }
\nonumber \\
&&
\
+\frac{1}{2}h^{2}\otimes \zeta \eta ^{2}\widehat{e_{-1,0}}e_{1,2}e^{-2\omega
\left( \zeta ,\eta \right) }+\frac{1}{2}\zeta ^{2}\eta
^{2}h^{2}e_{1,3}\otimes \widehat{e_{0,-1}}e_{1,3}e^{-\frac{1}{2}\sigma
\left( \zeta \right) -\frac{1}{2}\omega \left( \zeta ,\eta \right) }
\nonumber \\
&&
\
-\frac{1}{3}h^{3}\otimes \zeta \eta ^{3}\widehat{e_{-1,0}}e_{1,3}e^{-3\omega
\left( \zeta ,\eta \right) }+h\widehat{e_{-1,0}}\otimes \zeta \eta
e_{1,1}e^{-\frac{3}{2}\sigma \left( \zeta \right) +\frac{1}{2}\omega \left(
\zeta ,\eta \right) }
\nonumber \\
&&
\
-\frac{1}{2}\zeta \eta ^{2}h^{2}\widehat{e_{-1,0}}\otimes e_{1,2}e^{-\frac{1%
}{2}\sigma \left( \zeta \right) +\frac{1}{2}\omega \left( \zeta ,\eta
\right) }+\frac{1}{2}\zeta \eta ^{2}h^{2}e_{1,2}\otimes \widehat{e_{-1,0}}%
e^{-\frac{1}{2}\sigma \left( \zeta \right) -\frac{3}{2}\omega \left( \zeta
,\eta \right) }
\nonumber \\
&&
\
+\frac{1}{6}\zeta \eta ^{3}h^{3}e_{1,3}\otimes \widehat{e_{-1,0}}e^{-\frac{1%
}{2}\sigma \left( \zeta \right) -\frac{3}{2}\omega \left( \zeta ,\eta
\right) }+\frac{1}{6}\zeta \eta ^{3}h^{3}\widehat{e_{-1,0}}\otimes
e_{1,3}e^{-\frac{1}{2}\sigma \left( \zeta \right) -\frac{3}{2}\omega \left(
\zeta ,\eta \right) }
\nonumber \\
&&
\
+\zeta \eta he_{1,3}\otimes \widehat{e_{-1,-2}}e^{-\frac{1}{2}\sigma \left(
\zeta \right) +\frac{1}{2}\omega \left( \zeta ,\eta \right) }
\nonumber \\
&&
\
-\zeta ^{2}\eta he_{1,3}\otimes \widehat{e_{0,-1}}e_{1,2}e^{-\frac{1}{2}%
\sigma \left( \zeta \right) +\frac{1}{2}\omega \left( \zeta ,\eta \right) }+%
\frac{1}{2}\zeta ^{2}\left( e_{1,2}\right) ^{2}\otimes \widehat{e_{0,-1}}%
e^{-\sigma \left( \zeta \right) +\omega \left( \zeta ,\eta \right) }\, .
\cr &&
\end{eqnarray}
\el

We have finished the construction
  of the universal enveloping algebra $%
U_{C}(g_{2})$ twisted by the full chain of extended twists. The sequence of
factors (\ref{full-chain}) cannot be essentially enlarged due to the absence
of the Lie-Frobenius subalgebras in $g_{2}$ that nontrivially contain
${\mathfrak{b}}_{+}(g_{2})$
(see the discussion in Section~II). It is certainly possible to
perform the additional Abelian twist
 ${{{\cal F}}}_{A^{\prime }}=e^{\sigma \otimes
\omega }$, but in such a case   the carrier $\mathfrak{g}_{C}^{\#}$ is not
 changed.
 It appears that adding  the twist
   ${{{\cal F}}}_{A^{\prime }}$
   leads to the different realization of the same $%
\mathfrak{g}_{C}^{\#}$ in terms of the generators of $U_{C}(g_{2})$.

When such change of the realization of the carrier happens in the
intermediate steps of the quantization this can lead to interesting
results.
 We shall study this possibility in the next Subsection.

\subsection{General Form of Extended Twist for $g_{2}$:  Additional Abelian Twist}

Let us construct the quantization for the $r$-matrix (\ref{ws2.24b}). As it
was indicated in Section~III this can be performed by the twisting element
\begin{equation}\label{bolu5.25}
{{{\cal F}}}_{AE}\left( \rho \right) =e^{\rho h\otimes \sigma }e^{-e_{1,2}\otimes
e_{1,1}e^{-\frac{1}{2}\sigma }}e^{e_{1,3}\otimes e_{1,0}e^{-\frac{1}{2}%
\sigma }}e^{H\otimes \sigma }.
\end{equation}
Here the parameter $\rho $ is written explicitly because of its specific
role. Notice that for any fixed $\rho $ the element ${{{\cal F}}}_{AE}$
 depends on the  deformation parameter $\xi $
 (due to the $\xi$-dependence of $\sigma$ (see (\ref{blm5.22})).
Studying the parametrized set $\left\{
{{{\cal F}}}_{AE}\left( \xi ,\rho \right) \right\} $ we are dealing with the family of
twists indexed by the parameter $\rho $.
We shall demonstrate that contrary
to the case of the deformation parameter %(such as
 $\xi$ where we get equivalent deformed algebras
 $U_{{{{\cal F}}}_{AE}(\xi, \rho)} \sim U_{{{{\cal F}}}_{AE}(\xi', \rho)}$
 there are nonzero
values of  $\rho $ for which the twisted algebras are inequivalent $%
U_{{{{\cal F}}}_{AE}\left( \xi ,\rho \right) }\nsim U_{{{{\cal F}}}_{AE}\left( \xi ,\rho ^{\prime
}\right) }$.

In order to construct the twisted Hopf algebra $U_{{{{\cal F}}}_{AE}\left( \rho \right) }\left(
g_{2}\right) $ we use the coproducts $\Delta _{EJ}$ (5.18-5.21) of
 the full extended
twisting (5.3) obtained in Subsection~V.A and apply the transformation $%
e^{\rho \mathrm{ad}h\otimes \sigma }$ corresponding to the Abelian twist
factor ${{{\cal F}}}_{A}=e^{\rho h\otimes \sigma }$. Each nontrivial twisting factor
induces the transformation of the dual group and the new dual coordinates
are to be constructed. Omitting the intermediate steps we present the
results in terms of the new dual
$\mathfrak{g}_{AE}^{\#}$ -basis:
\begin{eqnarray}\label{bolu5.26}
\overline{H} &=&H+\rho h, \cr
\overline{e_{-2,-3}} &=&\widetilde{\widetilde{e_{-2,-3}}}+\rho ^{2}h^{2}, \cr
\overline{e_{-1,0}} &=&\widetilde{\widetilde{e_{-1,0}}}-\rho he_{1,3}, \cr
\overline{e_{-1,-1}} &=&\widetilde{\widetilde{e_{-1,-1}}}+\rho he_{1,2}
\end{eqnarray}
(the other $\mathfrak{g}_{E}^{\#}$ -coordinates $\widetilde{\widetilde{x}}$
remain unchanged).

According to the properties of the  twist ${\cal F}_A$ the costructure on its
carrier is conserved:
\begin{eqnarray}\label{bolu5.27}
\Delta _{AE}\left( h\right)  &=&h\otimes 1+1\otimes h,
\cr\cr
\Delta _{AE}\left( e_{2,3}\right)  &=&e_{2,3}\otimes e^{\sigma }+1\otimes
e_{2,3}\, .
\end{eqnarray}
The changes in $\Delta _{AE}\left( \overline{H}\right) $ and in the long
sequence $\Delta _{AE}\left( e_{1,l}\right) $ ($l=0,1,2,3$)
 are correlated with the root structure of the extensions:

\begin{eqnarray}\label{bolu5.28}
\Delta _{AE}\left( \overline{H}\right)  &=&\left( \overline{H}\right)
\otimes e^{-\sigma }+1\otimes \left( \overline{H}\right)
\cr
&&
-e_{1,3}\otimes e_{1,0}e^{\left( \rho \left( \alpha +3\beta \right) \left(
h\right) -\frac{3}{2}\right) \sigma }+e_{1,2}\otimes e_{1,1}e^{\left( \rho
\left( \alpha +2\beta \right) \left( h\right) -\frac{3}{2}\right) \sigma }
\cr
&=&
\left( \overline{H}\right) \otimes e^{-\sigma }+1\otimes \left( \overline{%
H}\right)
\cr
&&
 -e_{1,3}\otimes e_{1,0}e^{\frac{3}{2}\left( \rho -1\right) \sigma
}+e_{1,2}\otimes e_{1,1}e^{\frac{1}{2}\left( \rho -3\right) \sigma }\, .
\cr &&
\end{eqnarray}
\begin{eqnarray*}
\Delta _{AE}\left( e_{1,k}\right)  &=&e_{1,0}\otimes e^{\left( \frac{1}{2}%
+\rho \left( \alpha +k\beta \right) \left( h\right) \right) \sigma
}+e^{\sigma }\otimes e_{1,0},\quad k=0,1 \\
\Delta _{AE}\left( e_{1,m}\right)  &=&e_{1,0}\otimes e^{\left( -\frac{1}{2}%
+\rho \left( \alpha +m\beta \right) \left( h\right) \right) \sigma
}+1\otimes e_{1,0},\quad m=2,3
\end{eqnarray*}
On the plane orthogonal to the highest root we find the quasiprimitive
costructure:
\begin{eqnarray}
\Delta _{AE}\left( \widetilde{\widetilde{e_{0,1}}}\right)  &=&\widetilde{%
\widetilde{e_{0,1}}}\otimes e^{\rho \sigma }+1\otimes \widetilde{\widetilde{%
e_{0,1}}},  \notag \\
\Delta _{AE}\left( \widetilde{\widetilde{e_{0,-1}}}\right)  &=&\widetilde{%
\widetilde{e_{0,-1}}}\otimes e^{-\rho \sigma }+1\otimes \widetilde{%
\widetilde{e_{0,-1}}}\, .  \label{r-orthog}
\end{eqnarray}

In the coproducts for the other basic elements (belonging to
${\mathfrak{n}}_{-}$)do appear
 the
additional terms proportional to $\rho $
\bl
\begin{eqnarray}
\Delta _{AE}\left( \widetilde{\widetilde{e_{-1,-3}}}\right)  &=&\widetilde{%
\widetilde{e_{-1,-3}}}\otimes e^{-\frac{1}{2}\left( 1+3\rho \right) \sigma
}
\cr
&&
+1\otimes \widetilde{\widetilde{e_{-1,-3}}}+\left( 1-\rho \right) h\otimes
e_{1,0}e^{-\sigma }+\widetilde{\widetilde{e_{0,-1}}}\otimes
e_{1,1}e^{-\left( 1+\rho \right) \sigma }, \cr
&& \\
\Delta _{AE}\left( e_{-1,-2}\right)  &=&e_{-1,-2}\otimes e^{-\frac{1}{2}%
\left( 1+\rho \right) \sigma }+1\otimes e_{-1,-2}
\cr
&&
+\left( \rho -\frac{1}{3}%
\right) h\otimes e_{1,1}e^{-\sigma }-\widetilde{\widetilde{e_{0,1}}}\otimes
e_{1,0}e^{\left( \rho -1\right) \sigma }
\cr
&&+e_{1,2}\otimes \widetilde{\widetilde{e_{0,-1}}}e^{-\frac{1}{2}\sigma
\left( 1-\rho \right) },
\cr
&& \\
\Delta _{AE}\left( \overline{e_{-1,-1}}\right)  &=&\overline{e_{-1,-1}}%
\otimes e^{\frac{1}{2}\left( \rho -1\right) \sigma }+1\otimes \overline{%
e_{-1,-1}}
\cr
&&
+\left( \frac{1}{3}+\rho \right) e_{1,2}\otimes he^{\frac{1}{2}%
\left( \rho -1\right) \sigma }
\cr
&&-e_{1,3}\otimes \widetilde{\widetilde{e_{0,-1}}}e^{\frac{1}{2}\left( 3\rho
-1\right) \sigma }+\frac{4}{3}\widetilde{\widetilde{e_{0,1}}}\otimes
e_{1,1}e^{\left( \rho -1\right) \sigma },
\cr
&& \\
\Delta _{AE}\left( \overline{e_{-1,0}}\right)  &=&\overline{e_{-1,0}}\otimes
e^{\frac{1}{2}\left( 3\rho -1\right) \sigma }+1\otimes \overline{e_{-1,0}}%
+e_{1,2}\otimes \widetilde{\widetilde{e_{0,1}}}e^{-\frac{1}{2}\sigma \left(
1-\rho \right) }
\cr
&&
-\left( \rho +1\right) e_{1,3}\otimes he^{\frac{1}{2}\left(
3\rho -1\right) \sigma }\, ,
\cr &&
\\
\Delta _{AE}\left( \overline{e_{-2,-3}}\right) &=&\overline{e_{-2,-3}}%
\otimes e^{-\sigma }+1\otimes \overline{e_{-2,-3}}+2\rho h\otimes \overline{H%
}
\cr
&&-\overline{e_{-1,0}}\otimes e_{1,0}e^{\frac{3}{2}\left( \rho -1\right)
\sigma }-\overline{e_{-1,-1}}\otimes e_{1,1}e^{\frac{1}{2}\left( \rho
-3\right) \sigma }
\cr
&&
+e_{1,2}\otimes e_{-1,-2}e^{\frac{1}{2}\left( \rho
-1\right) \sigma }
\cr
&&
+e_{1,3}\otimes \widetilde{\widetilde{e_{-1,-3}}}e^{\frac{1}{2}\left(
3\rho -1\right) \sigma }-\frac{2}{3}\widetilde{\widetilde{e_{0,1}}}\otimes
\left( e_{1,1}\right) ^{2}e^{\left( \rho -2\right) \sigma }
 \cr
&&
+\frac{1}{2}\left( e_{1,2}\right) ^{2}\otimes \widetilde{\widetilde{%
e_{0,-1}}}e^{\left( \rho -1\right) \sigma }+2\rho he_{1,2}\otimes e_{1,1}e^{%
\frac{1}{2}\left( \rho -3\right) \sigma }\, .
\cr &&
\end{eqnarray}
\el
In the last expression the additional terms signify that the root vector $%
\overline{H}^{\ast }$ is no longer orthogonal to $h^{\ast }$ and the adjoint
operator $\mathrm{ad}\left( \overline{H}^{\ast }\right) $ transforms $h^{\ast }$
 into  $\overline{e_{-2,-3}}$.

In comparison with $\Delta _{E}\left( \widetilde{\widetilde{x}}\right) $ the
 changes of the form of  the coproducts $\Delta _{AE}$ are small but
   essential. We
immediately see five singular points $\rho =0,\pm 1,\pm \frac{1}{3}$. In
each of them one of the coproducts looses some terms and becomes closer to a
quasiprimitive. For example in the cases $\Delta _{AE_{+}}\left( \widetilde{%
\widetilde{e_{-1,-3}}}\right) $ and $\Delta _{AE_{+}}\left( \overline{%
e_{-1,0}}\right) $ if we use only the $sl(3)$-extension ${{{\cal F}}}_{+}=e^{e_{1,3}\otimes
e_{1,0}e^{-\frac{1}{2}\sigma }}$ then for $\rho =\pm 1$ these
coproducts become quasiprimitive and we obtain the possibility to perform
additional twistings with the carrier algebra nontrivially intersecting with
${\mathfrak{n}}_{-}$.
In the costructure $\Delta _{AE}$ the corresponding
enlargement of the carrier cannot be achieved, nevertheless the singular
points are also important. In particular we see that the standard case $%
\Delta _{EJ}$ corresponds to the singular point $\rho =0$ while in the
general situation the coproduct $\Delta _{AE\left( \rho \right) }\left(
\overline{e_{-2,-3}}\right) $ has additional components and
 we arrive at different
dual algebra $\mathfrak{g}_{AE}^{\#}\neq \mathfrak{g}_{E}^{\#}$.
 Notice that
in all the singular points the corresponding $r$-matrices differ only by
the value of the numerical parameter $\rho $ but the results of the
quantizations are different and refer to Lie-Poisson structures.

We have already seen that the coproducts $\Delta _{AE}\left( \widetilde{%
\widetilde{e_{0,\pm 1}}}\right) $ are now quasiprimitive.
 Consequently there
still exists the possibility
 to perform further twisting with the Jordanian
factor similar to ${{{\cal F}}}_{J^{\prime }}$
\begin{equation}
{{{\cal F}}}_{J^{\prime \prime }}=e^{h\otimes \omega ^{\prime }}
\end{equation}
with $\omega ^{\prime }=\ln \left( \widetilde{\widetilde{e_{0,1}}}e^{-\rho
\sigma }+e^{-\rho \sigma }\right) $. The twisting element
\begin{equation}
{{{\cal F}}}_{J^{\prime \prime }AE}=e^{h\otimes \omega ^{\prime }}e^{\rho h\otimes
\sigma }e^{-e_{1,2}\otimes e_{1,1}e^{-\frac{1}{2}\sigma }}e^{e_{1,3}\otimes
e_{1,0}e^{-\frac{1}{2}\sigma }}e^{H\otimes \sigma }
\end{equation}
is the solution of the twist equation. Here, contrary to the situation with
the full chain of extended twists, we do not obtain a parametrized family of
chain deformations.
It can be easily checked that (see also (\ref{full-chain}))

\begin{equation}
{{{\cal F}}}_{J^{\prime \prime }AE}={{{\cal F}}}_{C}=e^{h\otimes \omega }e^{-e_{1,2}\otimes
e_{1,1}e^{-\frac{1}{2}\sigma }}e^{e_{1,3}\otimes e_{1,0}e^{-\frac{1}{2}%
\sigma }}e^{H\otimes \sigma }\, .
\end{equation}
The dependence on $\rho $ cancels and we are again with the full chain
studied above in the Subsection~V.B. Thus we have obtained the following
result: there is the $\rho $-family of quantum algebras $U_{AE\left( \rho
\right) }\left( g_{2}\right) $ but the full chain quantization $U_{C}\left(
g_{2}\right) $ is unique.

To complete the analysis of the set of twist deformations for $U%
\left( g_{2}\right) $ let us consider the generators that can become
quasiprimitive (in general nonsimultaneously) after the action of
the full chain of twists $\mathcal{F}_{C}$. The number $q_{C}^{-}$
 of such generators is equal to the rank of $\frak{g}$. In our case
we have $%
q_{C}^{-}\left( g_{2}\right) =2$ and the corresponding generators
are $\widehat{e_{0,-1}}$ and $%
\widehat{e_{-1,0}}$. Both are quasiprimitive simultaneously but
cannot enlarge the space of $\frak{g}_{c}^{C}$ up to a
(quasi)Frobenius
subalgebra in $U\left( g_{2}\right) $. In the Borel
subalgebra $U_{C}\left(
{\mathfrak{b}}_{+}\left(g_{2}\right)\right) $ the number $q_{C}^{+}$ of
quasiprimitive generators is equal to the number of simultaneously
primitive.
For the full chains we have $q_{C}^{+}=r$ and in our case (as
we have seen above) these primitive
generators are $ \sigma $ and $\omega $. This certainly provides the
possibility to perform the additional Abelian twist but the corresponding
deformation will be equivalent to the redefinition of the Cartan elements
in the Jordanian twisting factors.

\section{Discussion and Outlook}

We have described the set $\left\{ U_{J},U_{E_{+}J},U_{EJ},U_{J^{^{\prime
}}EJ},U_{A\left( \rho \right) EJ}\right\} $ of quantized Lie-Poisson
structures that were constructed on the space $U\left( g_{2}\right) $ by
chains of twist deformations. The dual group coordinates ($\frak{g}^{\#}$%
-basis) obtained through the second classical limit procedure
  provide us with the possibility of writing down the explicit form
 of these Lie-Poisson structures. We have presented in this
 paper  the
 algebraic and coalgebraic formulae determined  by
 the exceptional Lie algebra $g_{2}$,
 in particular:

  -we get the additional (in comparison with the situation in $U\left(
sl(3)\right) $ and $U\left( so(5)\right) $) Hopf algebra $U_{EJ}$

 -the carrier space for the second Jordanian twist ${{{\cal F}}}_{J^{^{\prime
}}}$ is
deformed (the analogous result was found for $U_{E_{+}J}\left( so(5)\right)
$
 \cite{LLM,8basic})

 -the Hopf algebras $\left\{ U_{E_{+}J},U_{J^{^{\prime
}}EJ},U_{AEJ}\right\}$ in comparison with the analogous
quantizations of $U\left( sl(3)\right) $ and $U\left( so(5)\right)$ have
 more complicated costructure
 determined by the root system
 of $g_{2}$.

We have also found the peripheric twisted
 algebras in the set of Hopf algebras
 $\left\{ U_{A\left( \rho \right) EJ}\right\}$. The number $s_{C}$ of
  inequivalent algebras in this set
depends on the number $%
l_{C}^{n}$ of extended Jordanian factors in the chain
${{{\cal F}}}_{C}$.
  We conjecture the following relation
\begin{equation}
s_{C}=\sum_{i=1}^{l_{C}^{n}}\left( \dim \left( \frak{g}_{c}^{\left( i\right)
}\right) -1\right)\,  ,
\end{equation}
where $\frak{g}_{c}^{\left( i\right) }$ is the carrier subalgebra of the
link ${\cal F}^{\left( i\right) }\subset {{{\cal F}}}_{C}$. This gives three
inequivalent subsets for $%
U_{A\left( \rho \right) EJ}\left( sl(3)\right) $ as well as $U_{A\left( \rho
\right) EJ}\left( so(5)\right) $ and five for $U_{A\left( \rho \right)
EJ}\left( g_{2}\right) $. One of the algebras, $U_{A\left( 0\right)
EJ}\left( g_{2}\right) $, corresponds to the canonical extended twist and
four others are the analogues of the peripheric extended deformations
 \cite{8basic}
   in $U_{EJ}\left( sl(N)\right)$. At the same time we do not have peripheric
chains in $U_{{\cal F}}\left( g_{2}\right)$. As we have already stressed
above the full
chain deformation $U_{{\cal F}}(g_{2})$
 is invariant under the  rotation generated by additional Abelian twist.

\subsection*{Acknowledgments}
The paper is supported by
the Russian
 Foundations for Fundamental Research grants N 03-01-00837 (V.L.),
  RFBR-05-01-01086 (V.N.T.) and INTAS OPEN 03-51-3350 (V.N.T.).

\end{document}